\documentclass[twoside]{article}
\usepackage[accepted]{aistats2018}
\usepackage{amsfonts}
\usepackage{amsmath}
\usepackage{amsthm}
\usepackage{bm}
\usepackage{color}
\usepackage{dsfont} 
\usepackage[shortlabels]{enumitem}
\usepackage[authoryear,round]{natbib}
\usepackage{graphicx}
\usepackage[dvipsnames]{xcolor}
\usepackage{pdfpages}

\newcommand{\col}[2]{\bgroup\color{#1}#2\egroup}

\DeclareMathOperator*{\argmin}{argmin}
\renewcommand{\d}{\mathop{}\!\mathrm{d}}
\newcommand{\mnorm}[1]{{\vert\kern-0.25ex\vert\kern-0.25ex\vert #1 
    \vert\kern-0.25ex\vert\kern-0.25ex\vert}} 
    
\newtheorem{theorem}{Theorem}

\theoremstyle{remark}
   \newtheorem{remark}[theorem]{Remark}

  \newtheorem{example}[theorem]{Example}
\theoremstyle{definition}
\newtheorem{definition}[theorem]{Definition}

\newif\ifnocolor
\nocolortrue 

\usepackage[normalem]{ulem} 
\usepackage[colorinlistoftodos]{todonotes} 
\newcommand{\st}[1]{\ifmmode\text{\sout{\ensuremath{#1}}}\else\sout{#1}\fi} 
	\newcommand{\DELETE}[1]{\col{black!30}{\st{#1}}} 
	\newcommand{\DELETELONG}[1]{\col{black!30}{#1}} 

\ifnocolor
	\renewcommand{\DELETE}[1]{} 
	\renewcommand{\DELETELONG}[1]{} 
\fi

\begin{document}

%

%

\twocolumn[

\aistatstitle{Graphical Models for Non-Negative Data Using Generalized Score Matching}

\aistatsauthor{ Shiqing Yu \And Mathias Drton \And  Ali Shojaie}

\aistatsaddress{ University of Washington \And  University of Washington \And University of Washington } ]

\begin{abstract}
A common challenge in estimating parameters of probability density functions is the intractability of the normalizing constant. While in such cases maximum likelihood estimation may be implemented using numerical integration, the approach becomes computationally intensive. In contrast, the score matching method of \citet{hyv05} avoids direct calculation of the normalizing constant and yields closed-form estimates for exponential families of continuous distributions over $\mathbb{R}^m$.  \citet{hyv07} extended the approach to distributions supported on the non-negative orthant $\mathbb{R}_+^m$. In this paper, we give a generalized form of score matching for non-negative data that improves estimation efficiency. We also generalize the regularized score matching method of \citet{lin16} for non-negative Gaussian graphical models, with improved theoretical guarantees.
\end{abstract}

\section{INTRODUCTION}
\textit{Graphical models} \citep{lau96} characterize the relationships among random variables $(X_i)_{i\in V}$ indexed by the nodes of a graph $G=(V,E)$; here, $E \subseteq V\times V$ is the set of edges in $G$. When the graph $G$ is undirected, two variables $X_i$ and $X_j$ are required to be conditionally independent given all other $(X_k)_{k\in V\backslash\{i,j\}}$ if there is no edge between $i$ and $j$. The smallest graph $G$ such that this property holds is called the \emph{conditional independence graph} of the random vector $\boldsymbol{X}\equiv(X_i)_{i\in V}$. See \citet{drt17} for a more detailed introduction to these and other graphical models. 

Largely due to their tractability, Gaussian graphical models (GGMs)
have gained great popularity.   The conditional independence graph of a multivariate normal vector $\boldsymbol{X}\sim\mathcal{N}(\boldsymbol{\mu},\boldsymbol{\Sigma})$ is determined by the \emph{inverse covariance matrix} $\mathbf{K}\equiv\boldsymbol{\Sigma}^{-1}$, also known as the \emph{concentration matrix}. More specifically, $X_i$ and $X_j$ are conditionally independent given all other variables in $\boldsymbol{X}$ if and only if the $(i,j)$-th and the $(j,i)$-th entry of $\mathbf{K}$ are both zero. This simple relation underlies a rich literature on GGMs, including \citet{drt04}, \citet{mei06}, \citet{yua07} and \citet{fri08}, among others.

Recent work has provided tractable procedures also for non-Gaussian
graphical models. This includes Gaussian copula models
\citep{liu09,dob11,liu12}, Ising and other exponential family models \citep{rav10,che14,yang15}, as well as
semi- or non-parametric estimation techniques
\citep{fel13,voo13}.  In this paper, we focus on non-negative Gaussian
random variables, as recently considered by \citet{lin16} and
\citet{yu16}.

The \emph{probability density function} of a \emph{non-negative Gaussian} random vector $\boldsymbol{X}$ is proportional to that of the corresponding Gaussian vector, but restricted to the non-negative orthant. More specifically, let $\boldsymbol{\mu}$ and $\boldsymbol{\Sigma}$ be a mean vector and a covariance matrix for an $m$-variate random vector, respectively. Then $\boldsymbol{X}$ follows a truncated normal distribution with parameters $\boldsymbol{\mu}$ and $\boldsymbol{\Sigma}$ if it has density $\exp\left(-0.5(\boldsymbol{x}-\boldsymbol{\mu})^{\top}\mathbf{K}(\boldsymbol{x}-\boldsymbol{\mu})\right)$ on $\mathbb{R}_+^m\equiv[0,+\infty)^m$, where $\mathbf{K}\equiv\boldsymbol{\Sigma}^{-1}$ is the inverse covariance parameter. We denote this as $\boldsymbol{X}\sim\mathrm{TN}(\boldsymbol{\mu},\mathbf{K})$. The conditional independence graph of a truncated normal vector is determined by $\mathbf{K}\equiv[\kappa_{ij}]_{i,j}$ just as in the Gaussian case: $X_i$ and $X_j$ are conditionally independent given all other variables if $\kappa_{ij}=\kappa_{ji}=0$. 


Suppose $\boldsymbol{X}$ is a continuous random vector with distribution $P_0$, density $p_0$ with respect to Lebesgue measure, and support $\mathbb{R}^m$, so $p_{0}(\boldsymbol{x})\neq 0$ for all $\boldsymbol{x}\in\mathbb{R}^m$. Let $\mathcal{P}$ be a family of distributions with twice differentiable densities that we know only up to a (possibly intractable) normalizing constant. The score matching estimator of $p_0$ using $\mathcal{P}$ as a model is the minimizer of the expected squared $\ell_2$ distance between the gradients of $\log p_0$ and a log-density from $\mathcal{P}$. Formally, we minimize the loss from (\ref{eq_sm}) below. Although the loss depends on $p_0$, partial integration can be used to rewrite it in a form that can be approximated by averaging over the sample without knowing $p_0$. The key advantage of score matching is that the normalizing constant cancels from the gradient of log-densities. Furthermore, for exponential families, the loss is quadratic in the parameter of interest, making optimization straightforward. 

When dealing with distributions supported on a proper subset of $\mathbb{R}^m$, the partial integration arguments underlying the score matching estimator may fail due to discontinuities at the boundary of the support. To circumvent this problem, \citet{hyv07} introduced a \emph{modified score matching estimator} for data supported on $\mathbb{R}_+^m$ by minimizing a loss in which boundary effects are dampened by multiplying gradients elementwise with the identity functions $x_j$; see (\ref{eq_nn_sm}) below.
\citet{lin16} estimate truncated GGMs based on this modification, with an $\ell_1$ penalty on the entries of $\mathbf{K}$ added to the loss. In this paper, we show that elementwise multiplication with functions other than $x_j$ can lead to improved estimation accuracy in both simulations and theory. Following \citet{lin16}, we will then use the proposed generalized score matching framework to estimate the matrix $\mathbf{K}$. 

The rest of the paper is organized as follows. Section \ref{Score Matching} introduces score matching and our proposed \emph{generalized score matching}. In Section \ref{Exponential_Families}, we apply generalized score matching to exponential families, with univariate truncated Gaussian distributions as an example. \emph{Regularized generalized score matching} for graphical models is formulated in Section \ref{Regularized Generalized Score Matching}. Simulation results are given in Section \ref{Simulation Results}.

\subsection{Notation}

Subscripts are used to refer to entries in vectors and columns in matrices. Superscripts are used to refer to rows in matrices.
For example, when considering a matrix of observations $\mathbf{x}\in\mathbb{R}^{n\times m}$, each row being a sample of $m$ measurements/features for one observation/individual, $X_{j}^{(i)}$ is the $j$-th feature for the $i$-th observation. For a random vector $\boldsymbol{X}$, $X_j$ refers to its $j$-th component. 

The vectorization of a matrix $\mathbf{K}=[\kappa_{ij}]_{i,j}\in\mathbb{R}^{q\times r}$ is obtained by stacking its columns: 
\begin{multline*}
\mathrm{vec}(\mathbf{K})=(\kappa_{11},\ldots,\kappa_{q1},\kappa_{12},\ldots,\kappa_{q2},\\\ldots,\kappa_{1r},\ldots,\kappa_{qr})^{\top}.
\end{multline*}

For ${a}\geq 1$, the $\ell_{a}$-norm of a vector $\boldsymbol{v}\in\mathbb{R}^q$ is denoted 
$\|\boldsymbol{v}\|_{a}=(\sum_{j=1}^{m} |v_j|^q)^{1/{a}},$
and the $\ell_{\infty}$-norm is defined as $\|\boldsymbol{v}\|_{\infty}=\max_{j=1,\ldots,q}|v_j|$.
The $\ell_a$-$\ell_b$ operator norm for matrix $\mathbf{K}\in\mathbb{R}^{q\times r}$ is written as
$\mnorm{\mathbf{K}}_{a,b}\equiv\max_{\boldsymbol{x}\neq\boldsymbol{0}}\|\mathbf{K}\boldsymbol{x}\|_b/\|\boldsymbol{x}\|_a$
with shorthand notation $\mnorm{\mathbf{K}}_{a}\equiv\mnorm{\mathbf{K}}_{a,a}$. By definition, $\mnorm{\mathbf{K}}_{\infty}\equiv\max_{i=1.\ldots,q}\sum_{j=1}^{r}|\kappa_{ij}|$.
We write the Frobenius norm of $\mathbf{K}$ as
$\mnorm{\mathbf{K}}_{F}\equiv\|\mathrm{vec}(\mathbf{K})\|_{2}\equiv(\sum_{i=1}^q\sum_{j=1}^r\kappa_{ij}^2)^{1/2}$,
and its max norm
$\|\mathbf{K}\|_{\infty}\equiv\|\mathrm{vec}(\mathbf{K})\|_{\infty}\equiv\max_{i,j}|\kappa_{ij}|.$

For a scalar function $f$, we define $\partial_j f(\boldsymbol{x})$ as its partial derivative with respect to the $j$-th component evaluated at $x_j$, and $\partial_{jj}f(\boldsymbol{x})$ the corresponding second partial derivative. For vector-valued $\boldsymbol{f}:\mathbb{R}\to\mathbb{R}^m$, $\boldsymbol{f}(x)=(f_1(x),\ldots,f_m(x))^{\top}$, let $\boldsymbol{f}'(x)=(f_1'(x), \ldots, f_m'(x))^{\top}$ be the vector of derivatives and $\boldsymbol{f}''(x)$ likewise.

Throughout the paper, $\mathbf{1}_n$ refers to a vector of all $1$'s of length $n$. For $\boldsymbol{a}$, $\boldsymbol{b}\in\mathbb{R}^m$, $\boldsymbol{a}\circ\boldsymbol{b}\equiv(a_1b_1,\ldots,a_mb_m)^{\top}$. Moreover, when we speak of the ``density'' of a distribution, we mean its probability density function w.r.t.~Lebesgue measure. When it is clear from the context, $\mathbb{E}_0$ denotes the expectation under the true distribution.

\section{SCORE MATCHING}\label{Score Matching}
\subsection{Original Score Matching}
Suppose $\boldsymbol{X}$ is a random vector taking values in $\mathbb{R}^m$ with distribution $P_0$ and density $p_0$. Suppose $P_0\in\mathcal{P}$, a family of
distributions with twice differentiable densities supported on $\mathbb{R}^m$. 
The \emph{score matching loss} for $P\in\mathcal{P}$, with density $p$, is 
\begin{equation}\label{eq_sm}
J(P)=\int_{\mathbb{R}^m}p_0(\boldsymbol{x})\|\nabla\log p(\boldsymbol{x})-\nabla\log p_0(\boldsymbol{x})\|_2^2\d\boldsymbol{x}.
\end{equation}
The gradients in~(\ref{eq_sm}) can be thought of as
gradients with respect to a hypothetical location parameter, evaluated
at $\boldsymbol{0}$
\citep{hyv05}.
The loss $J(P)$ is minimized if and only if $P=P_0$, which forms the
basis for estimation of $P_0$.  Importantly, since the loss depends on
$p$ only through its log-gradient, it suffices to know $p$ up to a
normalizing constant. Under mild conditions, (\ref{eq_sm}) can be
rewritten as
\begin{multline}\label{eq_sm_eq}
J(P)=\int_{\mathbb{R}^m}p_0(\boldsymbol{x})\\\sum\limits_{j=1}^m\left[\partial_{jj}\log p(\boldsymbol{x})+\frac{\left(\partial_{j}\log p(\boldsymbol{x})\right)^2}{2}\right]\d \boldsymbol{x}
\end{multline}
plus a constant independent of $p$. Clearly, the integral in (\ref{eq_sm_eq}) can be approximated by its corresponding sample average without knowing the true density $p_0$, and can thus be used to estimate $p_0$.

\subsection{Score Matching for Non-Negative Data}
When the true density $p_0$ is only supported on a proper subset of
$\mathbb{R}^m$, the integration by parts underlying the equivalence of
(\ref{eq_sm}) and (\ref{eq_sm_eq}) may fail due to discontinuity at
the boundary. For distributions supported on the non-negative orthant $\mathbb{R}_+^m$, \citet{hyv07} addressed this
issue by instead minimizing the \emph{non-negative score matching
  loss}
\begin{multline}\label{eq_nn_sm}
J_+(P)=\int_{\mathbb{R}^m_+}p_0(\boldsymbol{x})\|\nabla\log p(\boldsymbol{x})\circ\boldsymbol{x}\\
-\nabla\log p_0(\boldsymbol{x})\circ\boldsymbol{x}\|_2^2\d\boldsymbol{x}.
\end{multline}
This loss can be motivated by considering gradients of the true and
model log-densities w.r.t.~a hypothetical scale parameter
\citep{hyv07}. Under regularity conditions, it can again be
rewritten as the expectation (under $P_0$) of a function independent
of $p_0$, thus allowing one to estimate $p_0$ by minimizing the
corresponding sample loss.

\subsection{Generalized Score Matching for Non-Negative Data}\label{Generalized Score Matching for Non-Negative Data}

We consider the following generalization of the non-negative score matching loss (\ref{eq_nn_sm}).
\begin{definition}\label{definition_GSM_loss}
  Suppose random vector $\boldsymbol{X}\in\mathbb{R}_+^m$ has true distribution $P_0$ with density $p_0$ that
  is twice differentiable and supported on $\mathbb{R}_+^m$.  Let
  $\mathcal{P}_+$ be the family of all distributions with twice
  differentiable densities supported on $\mathbb{R}_+^m$, and suppose
  $P_0\in\mathcal{P}_+$.  Let
  $h_1,\dots,h_m:\mathbb{R}_+\to\mathbb{R}_+$ be a.e.~positive functions
  that are differentiable almost everywhere, and set
  $\boldsymbol{h}(\boldsymbol{x})=(h_1(x_1),\dots,h_m(x_m))^{\top}$.
  For $P\in\mathcal{P}_+$ with density $p$, the \emph{generalized
    $\boldsymbol{h}$-score matching loss} is
\begin{multline}\label{eq_gsm}
J_{\boldsymbol{h}}(p)=\int_{\mathbb{R}_+^m}\frac{1}{2}p_0(\boldsymbol{x})\|\nabla\log p(\boldsymbol{x})\circ \boldsymbol{h}(\boldsymbol{x})^{1/2}\\
-\nabla\log p_0(\boldsymbol{x})\circ \boldsymbol{h}(\boldsymbol{x})^{1/2}\|_2^2\d\boldsymbol{x},
\end{multline}
where $\boldsymbol{h}^{1/2}(\boldsymbol{x})\equiv(h_1^{1/2}(x_1),\ldots,h_m^{1/2}(x_m))^{\top}$.
\end{definition}

Choosing all $h_j(x)=x^2$ recovers the loss from (\ref{eq_nn_sm}).  The key intuition for our generalized score matching is that we keep the $h_j$ increasing but instead focus on functions that are
bounded or grow rather slowly.  This will result in
reliable higher moments, leading to better practical performance and
improved theoretical guarantees.  We note that our approach could also
be presented in terms of transformations of data; compare to Section
11 in \citet{par12}. In particular, log-transforming positive data into all of $\mathbb{R}^m$ and then applying (\ref{eq_sm}) is equivalent to (\ref{eq_nn_sm}).

We will consider the following assumptions:

\begin{align*}
(\text{A1}) &\,\,\, p_0(\boldsymbol{x})h_j(x_j)\partial_j\log p(\boldsymbol{x})\to 0 \text{ as }x_j\nearrow+\infty\\
&\,\,\,\text{and as }x_j\searrow 0^+,\,\forall \boldsymbol{x}_{-j}\in\mathbb{R}_{+}^{m-1},\, \forall p\in\mathcal{P}_+,\\
(\text{A2}) &\,\, \,\mathbb{E}_{p_0}\|\nabla\log p(\boldsymbol{X})\circ \boldsymbol{h}^{1/2}(\boldsymbol{X})\|_2^2<+\infty, \\
&\,\,\,\mathbb{E}_{p_0}\|(\nabla\log p(\boldsymbol{X})\circ\boldsymbol{h}(\boldsymbol{X}))'\|_1<+\infty,\quad \forall p\in\mathcal{P}_+,
\end{align*}
where $\forall p\in\mathcal{P}_+$ is a shorthand for ``for all $p$ being the density of some $P\in\mathcal{P}_+$'', and the prime symbol denotes component-wise differentiation.

Assumption (A1) validates integration by parts and (A2) ensures the
loss to be finite. 
We note that (A1) and (A2) are easily satisfied when we consider exponential
families with $\lim_{x\searrow 0^+}h_j(x)=0$. 

The following theorem states that we can rewrite $J_{\boldsymbol{h}}$ as an expectation (under $P_0$) of a function that does not depend on $p_0$, similar to (\ref{eq_sm_eq}).
\begin{theorem}\label{theorem_GSM_loss_alt}
Under (A1) and (A2), 
\begin{align}
J_{\boldsymbol{h}}(p)&=C+\int_{\mathbb{R}_+^m}p_0\sum_{j=1}^m\left[h_j'(x_j)\partial_j (\log p(\boldsymbol{x}))\right.\label{eq_gsm_eq}\\
&\hspace{-0.3in}+\left.h_j(x_j)\partial_{jj}(\log p(\boldsymbol{x}))+\frac{1}{2}h_j(x_j)\left(\partial_j(\log p(\boldsymbol{x}))\right)^2\right]\d\boldsymbol{x},\nonumber
\end{align}
where $C$ is a constant independent of $p$.
\end{theorem}

Given a data matrix $\mathbf{x}\in\mathbb{R}^{n\times m}$ with rows
$\boldsymbol{X}^{(i)}$, we define the sample version of (\ref{eq_gsm_eq})
as
\begin{multline*}\label{eq_gsm_sample}
\hat{J}_{\boldsymbol{h}}(p)=\frac{1}{n}\sum_{i=1}^n\sum_{j=1}^m\left\{h_j'(X_j^{(i)})\partial_j (\log p(\boldsymbol{X}^{(i)}))+  \phantom{ \hspace{-3in}h_j(X_j^{(i)})\left[\partial_{jj}(\log p(\boldsymbol{X}^{(i)}))+\frac{1}{2}\left(\partial_j(\log p(\boldsymbol{X}^{(i)}))\right)^2\right]}\right.\\
\left.h_j(X_j^{(i)})\left[\partial_{jj}(\log p(\boldsymbol{X}^{(i)}))+\frac{1}{2}\left(\partial_j(\log p(\boldsymbol{X}^{(i)}))\right)^2\right]\right\}.
\end{multline*}
We first clarify estimation consistency, in analogy to Corollary 3 in
\citet{hyv05}.
\begin{theorem}\label{theorem_gsm}
  Consider a model
  $\{P_{\boldsymbol{\theta}}:\boldsymbol{\theta}\in\boldsymbol{\Theta}\}\subset\mathcal{P}_+$
  with parameter space $\boldsymbol{\Theta}$, and suppose that the
  true data-generating distribution
  $P_0\equiv P_{\boldsymbol{\theta}_0}\in\mathcal{P}_+$ with density
  $p_0\equiv p_{\boldsymbol{\theta}_0}$. Assume that
  $P_{\boldsymbol{\theta}}=P_0$ if and only if
  $\boldsymbol{\theta}=\boldsymbol{\theta}_0$. Then the generalized
  $\boldsymbol{h}$-score matching estimator $\hat{\boldsymbol{\theta}}$
  obtained by minimization of $\hat{J}_{\boldsymbol{h}}\color{black}{(p_{\boldsymbol{\theta}})}$ 
  over 
  $\boldsymbol{\Theta}$ converges in probability to $\boldsymbol{\theta}_0$ as
  the sample size $n$ goes to infinity.
\end{theorem}

\section{EXPONENTIAL FAMILIES}\label{Exponential_Families}
In this section, we study the case where $\{p_{\boldsymbol{\theta}}:\boldsymbol{\theta}\in\boldsymbol{\Theta}\}$ is an
exponential family comprising continuous distributions with support
$\mathbb{R}_+^m$.  More specifically, we consider densities that are indexed by the canonical parameter
$\boldsymbol{\theta}\in\mathbb{R}^{r}$ and have the form
\[\log
  p_{\boldsymbol{\theta}}(\boldsymbol{x})=\boldsymbol{\theta}^{\top}\boldsymbol{t}(\boldsymbol{x})-\psi(\boldsymbol{\theta})+b(\boldsymbol{x}),\quad\boldsymbol{x}\in\mathbb{R}_+^m.\]
It is not difficult to show that under assumptions (A1) and (A2) from
Section \ref{Generalized Score Matching for Non-Negative Data}, the
empirical generalized $\boldsymbol{h}$-score matching loss
$\hat{J}_{\boldsymbol{h}}$ above can be rewritten as
\begin{equation}\label{eq_gsm_exponential}
\hat{J}_{\boldsymbol{h}}(p_{\boldsymbol{\theta}})=\frac{1}{2}\boldsymbol{\theta}^{\top}\boldsymbol{\Gamma}(\mathbf{x})\boldsymbol{\theta}-\boldsymbol{g}(\mathbf{x})^{\top}\boldsymbol{\theta}+\mathrm{const},
\end{equation}
where $\boldsymbol{\Gamma}\in\mathbb{R}^{{r}^2}$ and $\boldsymbol{g}\in\mathbb{R}^{r}$ are sample averages of functions of the data matrix $\mathbf{x}$ only; the detailed expressions are omitted here.

Define
$\boldsymbol{\Gamma}_0\equiv\mathbb{E}_{p_0}[\boldsymbol{\Gamma}(\mathbf{x})]$,
$\boldsymbol{g}_0\equiv\mathbb{E}_{p_0}[\boldsymbol{g}(\mathbf{x})]$, and
$\boldsymbol{\Sigma}_0\equiv\mathbb{E}_{p_0}[(\boldsymbol{\Gamma}(\mathbf{x})\boldsymbol{\theta}_0-g(\mathbf{x}))(\boldsymbol{\Gamma}(\mathbf{x})\boldsymbol{\theta}_0-g(\mathbf{x}))^{\top}]$.

\begin{theorem}\label{thm_exponential}
  Suppose that
  \vspace{-0.35cm}
  \begin{itemize}
  \item[] \hspace{-.5cm}(C1) \ $\boldsymbol{\Gamma}$ is invertible almost
    surely, and  \\[-0.55cm]
  \item[] \hspace{-.5cm}(C2) \ $\boldsymbol{\Gamma}_0$, $\boldsymbol{\Gamma}_0^{-1}$, $\boldsymbol{g}_0$ and $\boldsymbol{\Sigma}_0$ exist and are finite entry-wise.
  \end{itemize}
  \vspace{-0.35cm}
  Then the minimizer of (\ref{eq_gsm_exponential}) is a.s.~unique with solution $\hat{\boldsymbol{\theta}}\equiv\boldsymbol{\Gamma}(\mathbf{x})^{-1}\boldsymbol{g}(\mathbf{x})$, and
\begin{align*}
\hat{\boldsymbol{\theta}}\to_{\text{a.s.}}\boldsymbol{\theta}_0\,\,\,\,\text{and}\,\,\,\,\sqrt{n}(\hat{\boldsymbol{\theta}}-\boldsymbol{\theta}_0)\to_d\mathcal{N}_{r}\left(\boldsymbol{0},\boldsymbol{\Gamma}_0^{-1}\boldsymbol{\Sigma}_0\boldsymbol{\Gamma}_0^{-1}\right).
\end{align*}
\end{theorem}
We note that (C1) holds if $h_j(X_j)>0$ a.e.~and
$[\partial_j
\boldsymbol{t}(\boldsymbol{X}^{(1)}),\dots,\partial_j\boldsymbol{t}(\boldsymbol{X}^{(n)})]\in\mathbb{R}^{{r}\times
  n}$ has rank ${r}$ a.e.~for some $j=1,\dots,m$.

In the following examples, we assume (A1)--(A2) and (C1)--(C2).
\begin{example}\label{ex_nn_mu}
Consider univariate ($m={r}=1$) truncated Gaussian distributions with unknown mean parameter $\mu$ and known variance parameter $\sigma^2$, so
\[p_{\mu}(x)\propto\exp\left(-\left(x-\mu\right)^2/\left(2\sigma^2\right)\right),\quad x\in\mathbb{R}_+.\]
Then, given i.i.d.~samples $X_1,\dots,X_n\sim p_{\mu_0}$, the generalized $h$-score matching estimator of $\mu$ is
\[\hat{\mu}_{h}\equiv\frac{\sum_{i=1}^nh(X_i)X_i-\sigma^2h'(X_i)}{\sum_{i=1}^n h(X_i)}.\] 
If $\lim_{x\searrow 0^+}h(x)=0$, $\lim_{x\nearrow+\infty}h^2(x)(x-\mu_{0})p_{\mu_0}(x)=0$ and the expectations are finite, \[\sqrt{n}(\hat{\mu}_{h}-\mu_0)\to_d\mathcal{N}\left(0,\frac{\mathbb{E}_{0}[\sigma^2 h^2(X)+\sigma^4 {h'}^{2}(X)]}{\mathbb{E}_{0}^2[h(X)]}\right).\]
\end{example}

\begin{example}\label{ex_nn_sigma}
Consider univariate truncated Gaussian distributions with known mean parameter $\mu$ and unknown variance parameter $\sigma^2>0$. Then, given i.i.d.~samples $X_1,\dots,X_n\sim p_{\sigma_0^2}$, the generalized $h$-score estimator of $\sigma^2$ is
\begin{align*}
&\hat{\sigma}^2_{h}\equiv\frac{\sum_{i=1}^nh(X_i)(X_i-\mu)^2}{\sum_{i=1}^n
  h(X_i)+h'(X_i)(X_i-\mu)}.
\end{align*}
If, in addition to the assumptions in Example \ref{ex_nn_mu},
$\lim_{x\nearrow+\infty}h^2(x)(x-\mu)^3p_{\sigma^2_0}(x)=0$, then
$\sqrt{n}(\hat{\sigma}^2_{h}-\sigma_0^2)\to_d\mathcal{N}(0,\tau^2)$ with
\begin{align*}
\tau^2\!\equiv\!\frac{2\sigma^6_{0}\mathbb{E}_{0}[h^2(X)(X-\mu)^2]+\sigma^8_{0}\mathbb{E}_{0}[{h'}^2(X)(X-\mu)^2]}{\mathbb{E}^2_{0}[h(X)(X-\mu)^2]}.
\end{align*}
When $\mu_{0}=0$, $h(x)\equiv 1$ also satisfies (A1)--(A2) and
(C1)--(C2), and the resulting estimator corresponds to the sample variance, which obtains the Cram\'er-Rao lower bound.
\end{example}

\begin{remark}
In the case of univariate truncated Gaussian, an intuitive explanation that using a bounded $h$ gives better results than \citet{hyv07} goes as follows. When $\mu\gg\sigma$, there is effectively no truncation to the Gaussian distribution, and our method automatically adapts to using low moments in (\ref{eq_gsm}), since a bounded and increasing $h(x)$ becomes almost constant as it gets close to its asymptote for $x$ large. When $h(x)$ becomes constant, we get back to the original score matching for distributions on $\mathbb{R}$. In other cases, the truncation effect is significant, and similar to \citet{hyv07}, our estimator uses higher moments accordingly.
\end{remark}

\begin{figure}[t]
\centering
\includegraphics[scale=0.29]{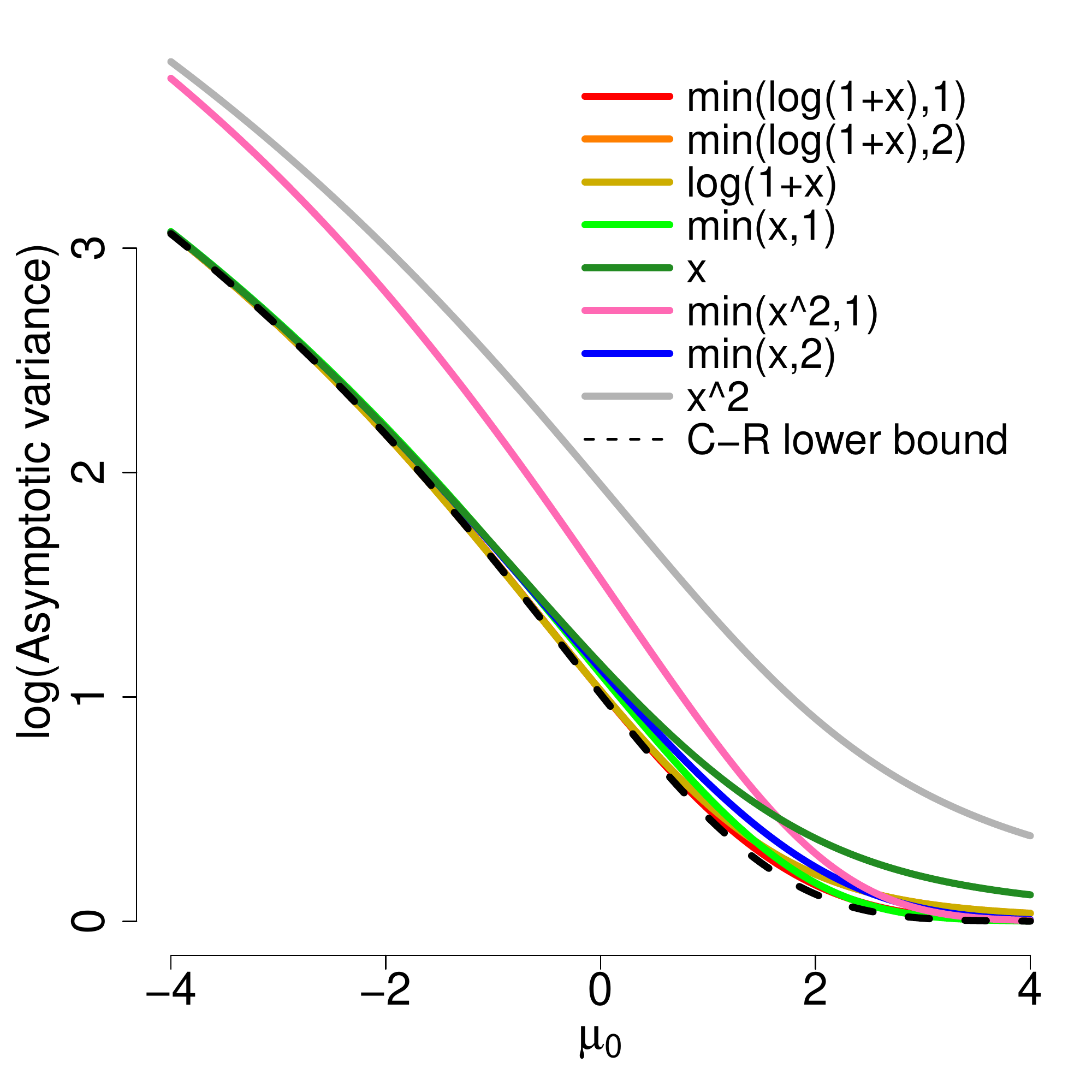}
\includegraphics[scale=0.29]{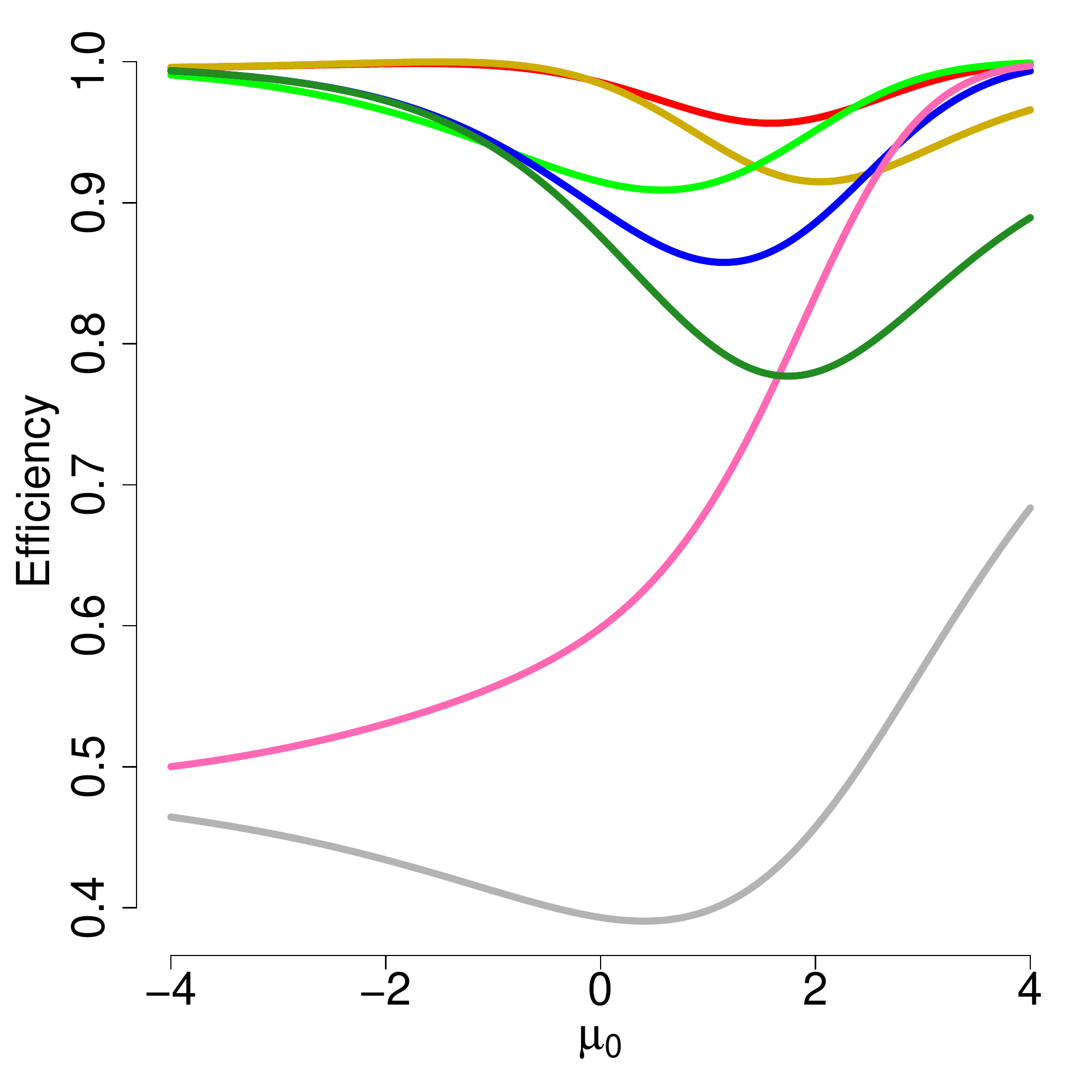}\caption{Log of asymptotic variance and efficiency w.r.t.~the Cram\'er-Rao bound for $\hat{\mu}_h$ ($\sigma^{2}=1$ known).}\label{plot_univariate_mu_var}
\end{figure}
Figure \ref{plot_univariate_mu_var} shows the theoretical asymptotic variance of $\hat{\mu}_{h}$ as given in Example \ref{ex_nn_mu}, with $\sigma=1$ known. Efficiency curves measured by the Cram\'er-Rao lower bound divided by the asymptotic variance are also shown. We see that two truncated versions of $\log(1+x)$ have asymptotic variance close to the Cram\'er-Rao bound. 
This asymptotic variance is also reflective of the variance for small finite samples.

\begin{figure}[t]
\centering
\includegraphics[scale=0.29]{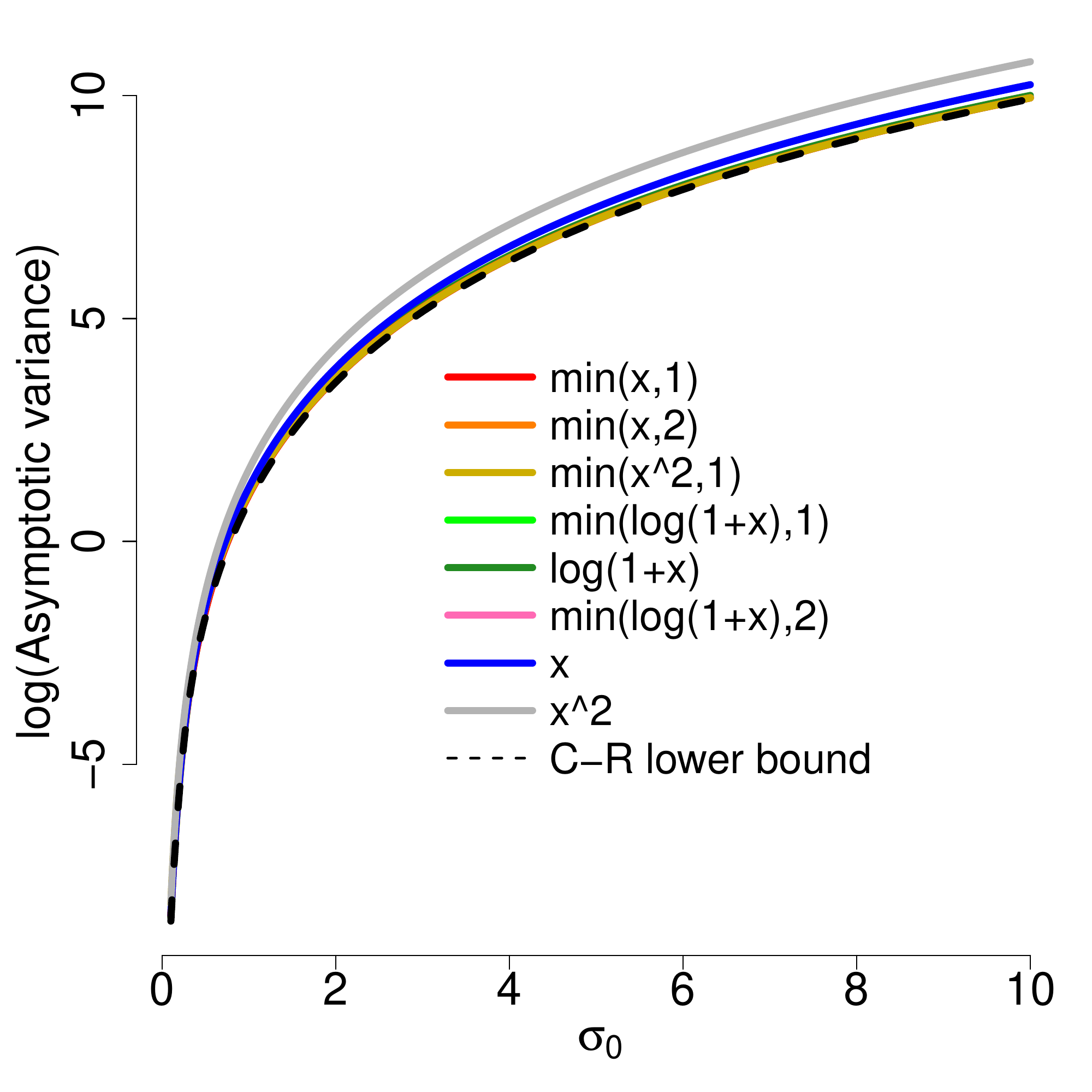}
\includegraphics[scale=0.29]{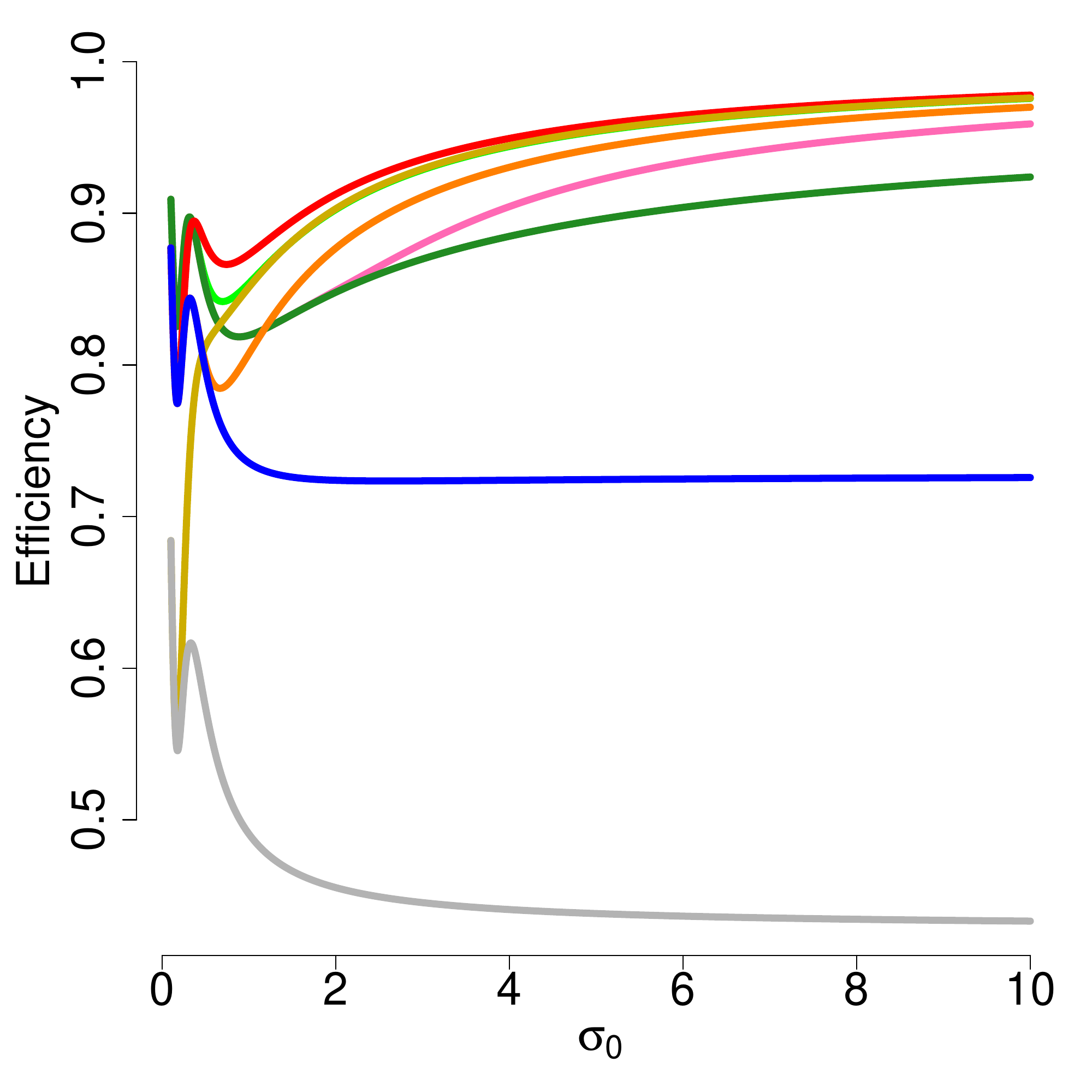}\caption{Log of asymptotic variance and efficiency w.r.t.~the Cram\'er-Rao bound for $\hat{\sigma}_h^2$ ($\mu=0.5$ known).}
\label{plot_univariate_sigma_var}
\end{figure}

Figure \ref{plot_univariate_sigma_var} is an analog of Figure \ref{plot_univariate_mu_var} for $\hat{\sigma}^2_{h}$ assuming $\mu=0.5$. For demonstration we choose a nonzero $\mu$ since when $\mu=0$ one can simply use the sample variance (degree of freedom unadjusted) which achieves the Cram\'er-Rao bound.

Here, the truncated versions of $x$ and $x^2$ have similar performance when $\sigma$ is not too small. In fact, when $\sigma$ is small, the truncation effect is small and one does not lose much by using the sample variance.

\section{REGULARIZED GENERALIZED SCORE MATCHING}\label{Regularized Generalized Score Matching}

We now turn to high-dimensional problems, where the number of parameters $r$ is larger than the sample size $n$. Targeting sparsity in the parameter $\boldsymbol{\theta}$, we consider $\ell_1$ regularization \citep{tib96}, which has also been used in graphical model estimation \citep{mei06,yua07,voo13}.

\begin{definition}
Assume the data matrix $\mathbf{x}\in\mathbb{R}^{n\times m}$ comprises
$n$ i.i.d.~samples from distribution $P_0$ with density $p_0$ that
belongs to an exponential family
$\{p_{\boldsymbol{\theta}}:\boldsymbol{\theta}\in\boldsymbol{\Theta}\}\subset\mathcal{P}_+$,
where 
$\boldsymbol{\Theta}\in\mathbb{R}^{r}$. Define the \emph{regularized generalized $\boldsymbol{h}$-score
  matching estimator} as
\begin{align}
\hat{\boldsymbol{\theta}}\,\in&\argmin\limits_{\boldsymbol{\theta}\in\boldsymbol{\Theta}}\hat{J}_{\boldsymbol{h},r}(\boldsymbol{\theta})\\
\equiv&\argmin\limits_{\boldsymbol{\theta}\in\boldsymbol{\Theta}}\frac{1}{2}\boldsymbol{\theta}^{\top}\boldsymbol{\Gamma}(\mathbf{x})\boldsymbol{\theta}-\boldsymbol{g}(\mathbf{x})^{\top}\boldsymbol{\theta}+\lambda\|\boldsymbol{\theta}\|_1\nonumber,
\end{align}
where $\lambda\geq 0$ is a tuning parameter, and $\boldsymbol{\Gamma}$ and $\boldsymbol{g}$ are from (\ref{eq_gsm_exponential}).
\end{definition}

Discussion of the general conditions for almost sure uniqueness of the solution is omitted here, but in practice we multiply the diagonals of $\boldsymbol{\Gamma}$ by a constant slightly larger than $1$ that ensures strict convexity and thus uniqueness of solution. Details about estimation consistency after this operation will be presented in future work. 
When the solution is unique, the solution path is \emph{piecewise linear}; compare \citet{lin16}.

\subsection{Truncated GGMs}

Using notation from the introduction, let $\boldsymbol{X}\sim\mathrm{TN}(\boldsymbol{\mu},\mathbf{K})$ be a truncated normal random vector with mean parameter $\boldsymbol{\mu}$ and inverse covariance/precision matrix parameter $\mathbf{K}$. Recall that the conditional independence graph for $\boldsymbol{X}$ corresponds to the support of $\mathbf{K}$, defined as $S\equiv S(\mathbf{K})\equiv\{(i,j):\kappa_{ij}\neq 0\}$. This support is our target of estimation.

\subsection{Truncated Centered GGMs}\label{Truncated Centered GGMs}
Consider the case where the mean parameter is zero, i.e., $\boldsymbol{\mu}\equiv\boldsymbol{0}$, and we want to estimate the inverse covariance matrix $\mathbf{K_{}}\in\mathbb{R}^{m^2}$. Assume that for all $j$ there exist constants $M_j$ and $M_j'$ that bound $h_j$ and its derivative $h_j'$ a.e. Assume further that $h_j(x)>0$ a.e, $\lim_{x\searrow 0^+}h_j(x)=0$, and $h_j'(x)\geq 0$. Boundedness here is for ease of proof in the main theorems; reasonable choices of unbounded $h$ are also valid. Then, (A1)--(A2) are satisfied, and the loss can be written as 
\begin{multline}\label{eq_loss_centered}
\hat{J}_{r+}(\mathbf{K})=\tfrac{1}{2}\mathrm{vec}(\mathbf{K})^{\top}\boldsymbol{\Gamma}(\mathbf{x})\mathrm{vec}(\mathbf{K})\\
-\boldsymbol{g}(\mathbf{x})^{\top}\mathrm{vec}(\mathbf{K})+\lambda\|\mathbf{K}\|_1,
\end{multline}
with the $j^{\mathrm{th}}$ block of the $m^2\times m^2$ block-diagonal matrix $\boldsymbol{\Gamma}(\mathbf{x})$ being
\begin{equation}\label{eq_centered_gamma}
n^{-1}\mathbf{x}^{\top}\mathrm{diag}(\boldsymbol{h}_j(\boldsymbol{X}_j))\mathbf{x},\nonumber
\end{equation}
where $\boldsymbol{h}_j(\boldsymbol{X}_j)\equiv[h_j(X_j^{(1)}),\ldots,h_j(X_{j}^{(n)})]^{\top}$, $\mathrm{diag}(c_1,\ldots,c_n)$ denotes a diagonal matrix with diagonal entries $c_1,\ldots,c_n$, and
\begin{alignat*}{3}
\boldsymbol{g}(\mathbf{x})&\equiv\mathrm{vec}(\mathbf{u})+\mathrm{vec}(\mathrm{diag}(\boldsymbol{V})), \,&&\,\,\mathbf{u}\equiv n^{-1}\mathbf{h'}(\mathbf{x})^{\top}\mathbf{x}, \\
\boldsymbol{V}&\equiv n^{-1}\mathbf{h}(\mathbf{x})^{\top}\mathbf{1}_n,\,&&\,\,\mathbf{h}(\mathbf{x})\equiv[h_j(X_{j}^{(i)})]_{i,j}, \\
\mathbf{h'}(\mathbf{x})&\equiv[h'_j(X_{j}^{(i)})]_{i,j},\,&&\,\,\mathbf{1}_n=[1,\ldots,1]^{\top}.
\end{alignat*}
The \emph{regularized generalized $\boldsymbol{h}$-score matching estimator} of $\mathbf{K}$ in the truncated centered GGM is 
\begin{equation}\label{eq_Khat_centered}
\hspace{-0.1in}\hat{\mathbf{K}}\equiv\argmin_{\substack{\mathbf{K}\in\mathbb{R}^{m^2},\mathbf{K}=\mathbf{K}^\top}}\hat{J}_{r+}(\mathbf{K}),
\end{equation}
where $\boldsymbol{\Gamma}(\mathbf{x})$ and $\boldsymbol{g}(\mathbf{x})$ are defined above.
\begin{definition}\label{def_constants_centered}
For true inverse covariance matrix $\mathbf{K}_0$, let $\boldsymbol{\Gamma}_0\equiv\mathbb{E}_{0}\boldsymbol{\Gamma}(\mathbf{x})$ and $\boldsymbol{g}_0\equiv\mathbb{E}_{0}\boldsymbol{g}(\mathbf{x})$. Denote the support of a precision matrix $\mathbf{K}$ as $S\equiv S(\mathbf{K})\equiv\{(i,j):\kappa_{ij}\neq 0\}.$ Write the true support of $\mathbf{K}_0$ as $S_0=S(\mathbf{K}_0)$. Suppose the maximum number of non-zero entries in rows of $\mathbf{K}_0$ is $d_{\mathbf{K}_0}$. Let $\mathbf{\Gamma}_{SS}$ be the entries of $\mathbf{\Gamma}\in\mathbb{R}^{m^2\times m^2}$ corresponding to edges in $S$. Define
\[c_{\boldsymbol{\Gamma}_0}\equiv\mnorm{(\boldsymbol{\Gamma}_{0,S_0S_0})^{-1}}_{\infty,\infty},\quad c_{\boldsymbol{K}_0}\equiv\mnorm{\mathbf{K}_0}_{\infty,\infty}.\] We say the irrepresentability condition holds for $\boldsymbol{\Gamma}_{0}$ if there exists an $\alpha\in(0,1]$ such that
\begin{equation}\label{irrepresentability}
\mnorm{\boldsymbol{\Gamma}_{0,S_0^cS_0}(\boldsymbol{\Gamma}_{0,S_0S_0})^{-1}}_{\infty,\infty}\leq (1-\alpha).
\end{equation}
\end{definition}

\begin{theorem}\label{thm_1}
Suppose $\boldsymbol{X}\sim\mathrm{TN}(\boldsymbol{0},\mathbf{K}_0)$ and $\boldsymbol{h}$ is as discussed in the opening paragraph of this section. Suppose further that $\boldsymbol{\Gamma}_{0,S_0S_0}$ is invertible and satisfies the irrepresentability condition (\ref{irrepresentability}) with $\alpha\in(0,1]$. Let $\tau>3$. If the sample size and the regularization parameter satisfy
\begin{align}
n&\geq\mathcal{O}\left(d_{\mathbf{K}_0}^2\tau\log m\max\left\{\frac{c_{\boldsymbol{\Gamma}_0}^2c_{\boldsymbol{X}}^4}{\alpha^2},1\right\}\right),\\
\lambda&>\mathcal{O}\left[(c_{\mathbf{K}_0}c_{\boldsymbol{X}}^2+c_{\boldsymbol{X}}+1)\left(
\sqrt{\frac{\tau\log m}{n}}+\frac{\tau\log m}{n}\right)\right],
\end{align}
where $c_{\boldsymbol{X}}^{}\equiv2\max_j\left(2\sqrt{(\mathbf{K}_0^{-1})_{jj}}+\sqrt{e}\mathbb{E}_{0} X_j\right)$, then the following statements hold with probability $1-m^{3-\tau}$:
\begin{enumerate}[(a)]
\item The regularized generalized $\boldsymbol{h}$-score matching estimator $\hat{\mathbf{K}}$ defined in (\ref{eq_Khat_centered}) is unique, has its support included in the true support, $\hat{S}\equiv S(\hat{\mathbf{K}})\subseteq S_0$, and
\begin{align*}
\|\hat{\mathbf{K}}-\mathbf{K}_0\|_{\infty}&<\frac{c_{\boldsymbol{\Gamma}_0}}{2-\alpha}\lambda,\\
\mnorm{\hat{\mathbf{K}}-\mathbf{K}_0}_{F}&\leq\frac{c_{\boldsymbol{\Gamma}_0}}{2-\alpha}\lambda\sqrt{|S_0|},\\
\mnorm{\hat{\mathbf{K}}-\mathbf{K}_0}_{2}&\leq\frac{c_{\boldsymbol{\Gamma}_0}}{2-\alpha}\lambda\min(\sqrt{|S_0|},d_{\mathbf{K}_0}).
\end{align*}
\item Moreover, if
\[\min_{j,k}|\kappa_{0,jk}|>\frac{c_{\boldsymbol{\Gamma}_0}}{2-\alpha}\lambda,\]
then $\hat{S}=S_0$ and $\mathrm{sign}(\hat{\kappa}_{jk})=\mathrm{sign}(\kappa_{0,jk})$ for all $(j,k)\in S_0$.
\end{enumerate}
\end{theorem}
The theorem is proved in the supplement.
A key ingredient of the proof is a tail bound on $\|\boldsymbol{\Gamma}-\boldsymbol{\Gamma}_0\|_{\infty}$, which is composed of products of the $X_j^{(i)}$'s. In \citet{lin16}, the products are up to fourth moments. Using a bounded $\boldsymbol{h}$ our products automatically calibrate to a quadratic polynomial when the observed values are large, and resort to higher moments only when they are small. Using this scaling, we obtain improved bounds and convergence rates, underscored in the new requirement on the sample size $n$, which should be compared to $n>\mathcal{O}((\log m^{\tau})^8)$ in \citet{lin16}. 

\subsection{Truncated Non-centered GGMs}\label{Truncated Non-centered GGMs}
Suppose now $\boldsymbol{X}\sim\mathrm{TN}(\boldsymbol{\mu}_0,\mathbf{K}_0)$ with both $\boldsymbol{\mu}_0$ and $\mathbf{K}_0$ unknown. While our main focus is still on the inverse covariance parameter $\mathbf{K}$, we now also have to estimate a mean parameter $\boldsymbol{\mu}$. Instead, we estimate the canonical parameters $\mathbf{K}_{}$ and $\boldsymbol{\eta}_{}\equiv\mathbf{K}_{}\boldsymbol{\mu}$. Concatenating $\boldsymbol{\Xi}_{}\equiv[\mathbf{K}_{},\boldsymbol{\eta}_{}]$, the corresponding $\boldsymbol{h}$-score matching loss has a similar form to (\ref{eq_loss_centered}) and (\ref{eq_Khat_centered}) with $\mathbf{K}$ replaced by $\boldsymbol{\Xi}$, and different $\boldsymbol{\Gamma}$ and $\boldsymbol{g}$. As a corollary of the centered case, we have an analogous bound on the error in the resulting estimator $\hat{\boldsymbol{\Xi}}$; we omit the details here. We note, however, that we can have different tuning penalty parameters $\lambda_{\mathbf{K}}$ and $\lambda_{\boldsymbol{\eta}}$ for $\mathbf{K}$ and $\boldsymbol{\eta}$, respectively, as long as their ratio is fixed, since we can scale the $\boldsymbol{\eta}$ parameter by the ratio accordingly. To avoid picking two tuning parameters, one may also choose to remove the penalty on $\boldsymbol{\eta}$ altogether by profiling out $\boldsymbol{\eta}$. We leave a detailed analysis of the profiled estimator to future research.

\subsection{Tuning Parameter Selection}\label{Parameter Tuning}
By treating the loss as the mean negative log-likelihood, we may use the extended Bayesian information Criterion (eBIC) to choose the tuning parameter \citep{che08,foy10}. Let $\hat{S}^{\lambda}\equiv\{(i,j):\hat{\kappa}_{ij}^{\lambda}\neq 0,i<j\}$, where $\hat{\mathbf{K}}^{\lambda}$ be the estimate associated with tuning parameter $\lambda$. The eBIC is then
\begin{multline*}
\mathrm{eBIC}(\lambda)=-n\mathrm{vec}(\hat{\mathbf{K}})^{\top}\boldsymbol{\Gamma}(\mathbf{x})\mathrm{vec}(\hat{\mathbf{K}})\\
+2n\boldsymbol{g}(\mathbf{x})^{\top}\mathrm{vec}(\hat{\mathbf{K}})+|\hat{S}^{\lambda}|\log n+2\log\left(\begin{matrix}p(p-1)/2 \\ |\hat{S}^{\lambda}|\end{matrix}\right),
\end{multline*}
where $\hat{\mathbf{K}}$ can be either the original estimate associated with $\lambda$, or a refitted solution obtained by restricting the support to $\hat{S}^{\lambda}$.

\section{NUMERICAL EXPERIMENTS}\label{Simulation Results}

We present simulation results for non-negative GGM estimators with different choices of $\boldsymbol{h}$. We use a common $h$ for all columns of the data matrix $\mathbf{x}$. Specifically, we consider functions such as $h_j(x)=x$ and $h_j(x)=\log(1+x)$ as well as truncations of these functions. In addition, we try MCP \citep{fan01} and SCAD penalty-like \citep{zha10} functions.

\subsection{Implementation}
We use a coordinate-descent method analogous to Algorithm 2 in \citet{lin16}, where in each step we update each element of $\hat{\mathbf{K}}$ based on the other entries from the previous steps, while maintaining symmetry. Warm starts using the solution from the previous $\lambda$, as well as lasso-type strong screening rules \citep{tib12} are used for speedups. In our simulations below we always scaled the data matrix by column $\ell_2$ norms before proceeding to estimation.

\subsection{Truncated Centered GGMs}\label{Truncated Centered Gaussian Graphical Models}

For data from a truncated centered Gaussian distribution, we compared our truncated centered Gaussian estimator (\ref{eq_Khat_centered}) with various choices of $h$, to \emph{SpaCE JAM} (SJ, \citealp{voo13}), which estimates graphs without assuming a specific form of distribution, a pseudo-likelihood method \emph{SPACE} \citep{pen09} with \emph{CONCORD} reformulation \citep{kha15}, \emph{graphical lasso} \citep{yua07,fri08}, {the} \emph{neighborhood selection} estimator (NS) in \citet{mei06}, and finally \emph{nonparanormal SKEPTIC} \citep{liu12} with Kendall's $\tau$. While we ran all these competitors, only the top performing methods are explicitly shown in our reported results. Recall that the choice of $h(x)=x^2$ corresponds to the estimator in \citet{lin16}, using score matching from \citet{hyv07}.
\begin{figure}[t]
\includegraphics[width=0.43\textwidth,scale=0.43]{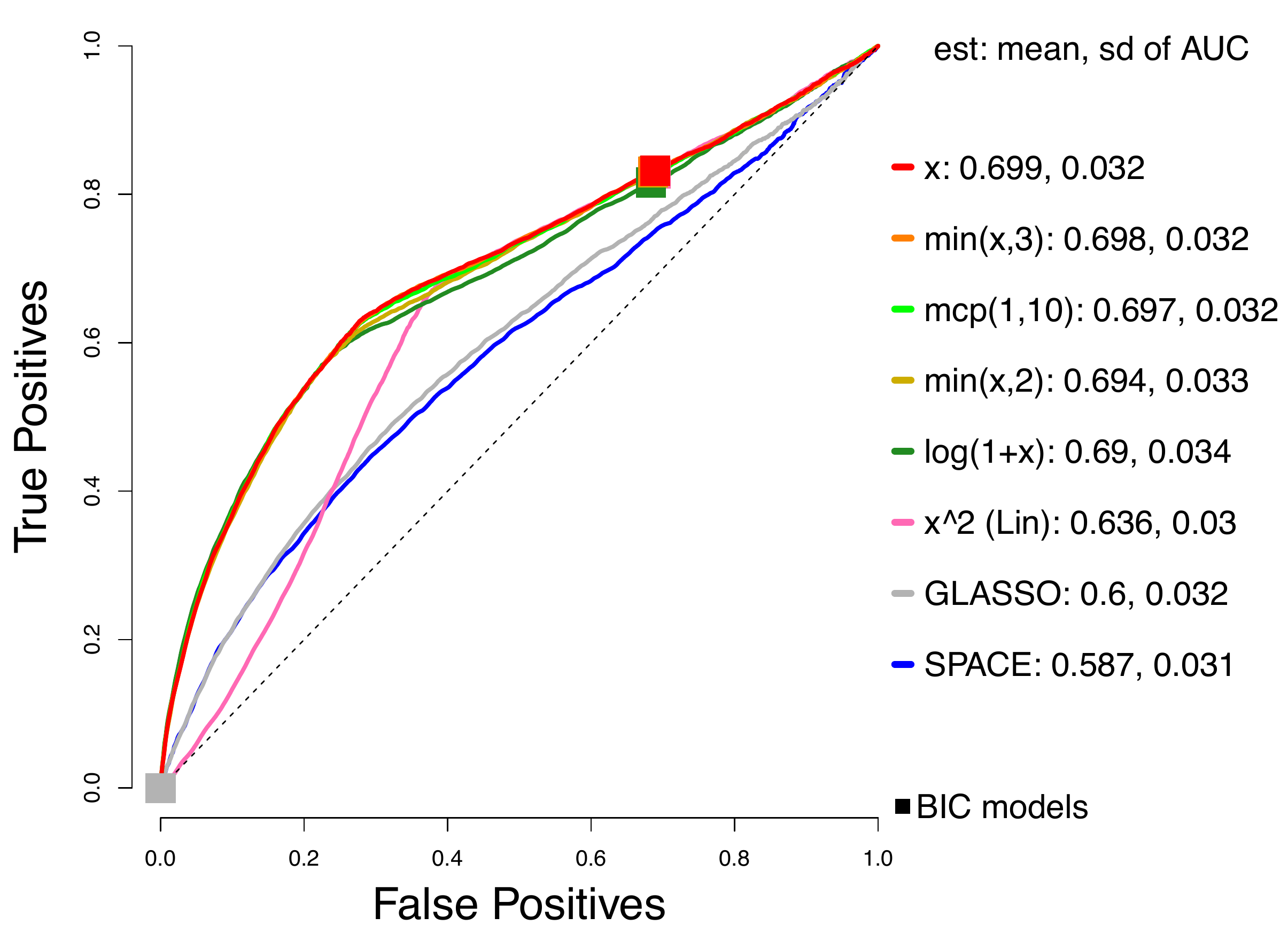}\caption{Average ROC curves of our estimator with various choices of $h$, compared to SPACE and GLASSO, for the \emph{truncated centered Gaussian} case. $n=80$, $m=100$. Squares indicate average true positive rate (TPR) and false positive rate (FPR) of models picked by eBIC (with refitting) for the estimator in the same color.  eBIC is introduced in Section \ref{Parameter Tuning}.}
\label{plot_centered_80}
\end{figure}

\begin{figure}[t]
\includegraphics[width=0.43\textwidth,scale=0.43]{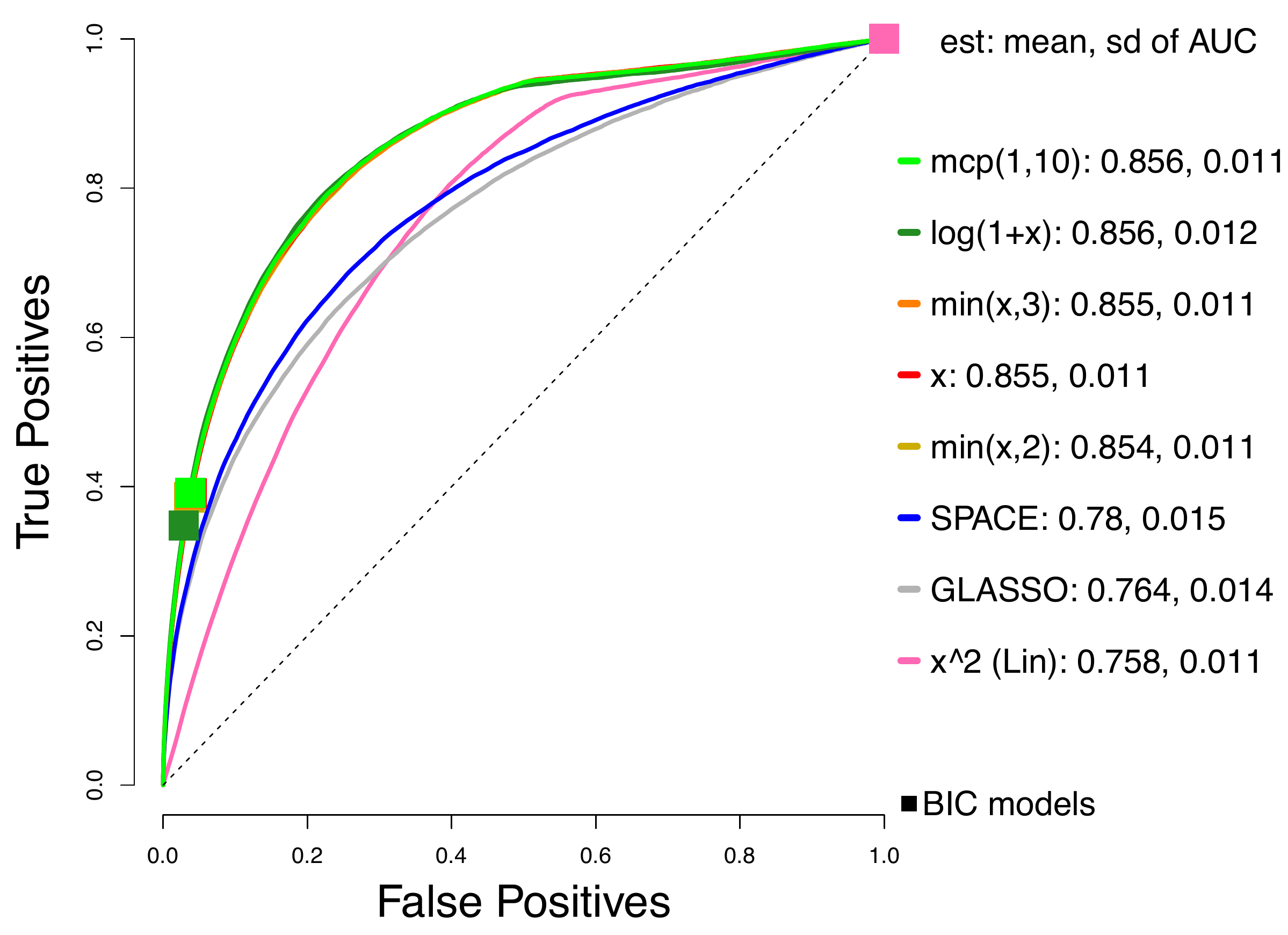}\caption{Same setting as in Figure \ref{plot_centered_80}, with $n=1000$.}
\label{plot_centered_1000}
\end{figure}

A few representative ROC (\emph{receiver operating characteristic}) curves for edge recovery are plotted in Figures \ref{plot_centered_80}  and \ref{plot_centered_1000}, using $m=100$ and $n=80$ or $n=1000$, respectively. Each curve corresponds to the average of 50 ROCs obtained from estimation of $\mathbf{K}_{}$ from $\mathbf{x}$ generated using 5 different true precision matrices $\mathbf{K}_0$, each with 10 trials. The averaging method is mean AUC-preserving and is introduced as \emph{vertical averaging} and outlined in Algorithm 3 in \citet{faw06}. The construction of $\mathbf{K}_0$ is the same as in Section 4.2 of \citet{lin16}: a graph with $m=100$ nodes with 10 disconnected subgraphs containing the same number of nodes, i.e.~$\mathbf{K}_0$ is block-diagonal. In each sub-matrix, we generate each lower triangular element to be $0$ with probability $\pi\in(0,1)$, and from a uniform distribution on interval $[0.5,1]$ with probability $1-\pi$. The upper triangular elements are set accordingly by symmetry. The diagonal elements of $\mathbf{K}_0$ are chosen to be a common positive value so that the minimum eigenvalue of $\mathbf{K}_0$ is $0.1$. We choose $\pi=0.2$ for $n=80$ and $\pi=0.8$ for $n=1000$.

For clarity, we only plot some top-performing representatives of the functions we considered. However, all of the alternative functions $h$ we considered perform better than $h(x)=x^2$ from \citet{hyv07} and \citet{lin16}. 

\subsection{Truncated Non-Centered GGMs}

Next we generate data from a truncated non-centered Gaussian distribution with both parameters $\boldsymbol{\mu}_{}$ and $\mathbf{K}_{}$ unknown. Consider $\hat{\mathbf{K}}$ as part of the estimated $\hat{\boldsymbol{\Xi}}$ as discussed in Section \ref{Truncated Non-centered GGMs}. In each trial we form the true $\mathbf{K}_0$ as in Section \ref{Truncated Centered Gaussian Graphical Models}, and we generate each component of $\boldsymbol{\mu}_0$ independently from the normal distribution with mean $0$ and standard deviation $0.5$.

\begin{figure}[t]
\includegraphics[width=0.43\textwidth,scale=0.43]{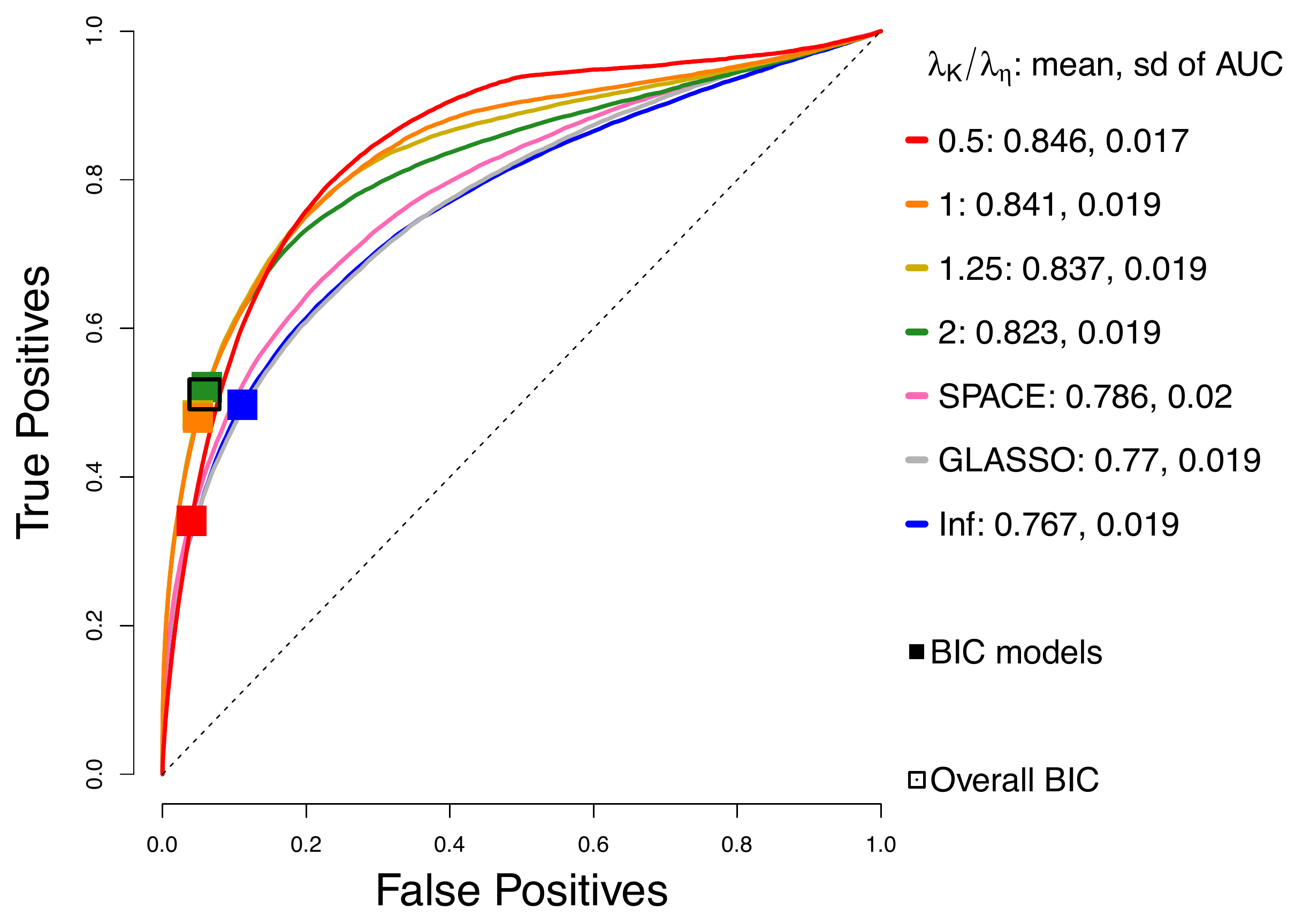}\caption{Performance of the non-centered estimator with $h(x)=\min(x,3)$; each curve corresponds to a different choice of fixed $\lambda_{\mathbf{K}}/\lambda_{\boldsymbol{\eta}}$. $n=1000$, $m=100$.}
\label{plot_lr_1000_min}
\end{figure}

As discussed in Section \ref{Truncated Non-centered GGMs}, we assume the ratio of the tuning parameters for $\mathbf{K}$ and $\boldsymbol{\eta}$ to be fixed. Shown in Figure \ref{plot_lr_1000_min} are average ROC curves (over 50 trials as in Section \ref{Truncated Centered Gaussian Graphical Models}) for truncated non-centered GGM estimators with $h(x)=\min(x,3)$; each curve corresponds to a different ratio ${\lambda_{\mathbf{K}}}/{\lambda_{\boldsymbol{\eta}}}$, where ``Inf'' indicates $\lambda_{\boldsymbol{\eta}}\equiv 0$. Here, $n=1000$ and $m=100$. 

Clearly, as the ratio increases, the performance improves, and after a certain threshold it deteriorates. The AUC for the profiled estimator with $\lambda_{\boldsymbol{\eta}}=0$ is among the worst, so there indeed is a lot to be gained from tuning an extra tuning parameter, although there is a tradeoff between time and performance. 

\begin{figure}[t]
\includegraphics[width=0.43\textwidth,scale=0.43]{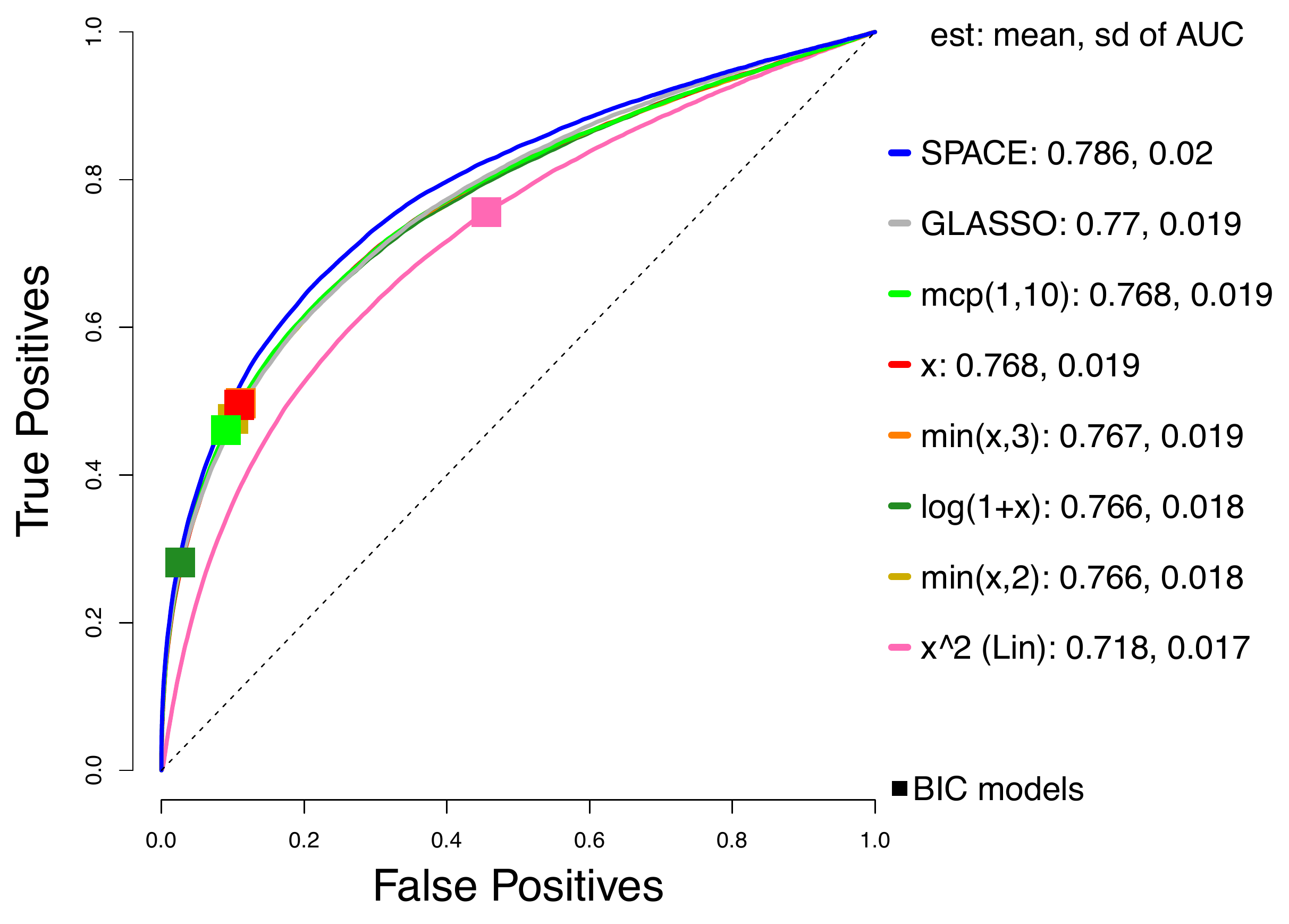}\caption{Performance of the non-centered profiled estimators using various $h$, $n=1000$, $m=100$. 
}
\label{plot_profiled_1000}
\end{figure}

In Figure \ref{plot_profiled_1000} we compare the performance of the profiled estimator with different $h$, to SPACE and GLASSO, each with 50 trials as before. It can be seen that even without tuning the extra parameter, the estimators, except for $h(x)=x^2$, still work as well as SPACE and GLASSO, and outperform \citet{lin16}.

\section{DISCUSSION}

In this paper we proposed a generalized version of the score matching estimator of \citet{hyv07} which avoids the calculation of normalizing constants. For estimation of the canonical parameters of exponential families, our generalized loss retains the nice property of being quadratic in the parameters. Our estimator offers improved estimation properties through various scalar or vector-valued choices of function $\boldsymbol{h}$.

For high-dimensional exponential family graphical models, following the work of \citet{mei06}, \citet{yua07} and \citet{lin16}, we add an $\ell_1$ penalty to the generalized score matching loss, giving a solution that is almost surely unique under regularity conditions and has a piecewise linear solution path.

In the case of multivariate truncated Gaussian distribution, where the conditional independence graph is given by the inverse covariance parameter, the sample size required for the consistency of our method is $\Omega(d^2\log m)$, where $m$ is the dimension and $d$ is the maximum node degree in the corresponding independence graph. This matches the rates for GGMs in \citet{rav11} and \citet{lin16}, and lasso with linear regression \citep{buh11}.

A potential problem for future work would be adaptive choice of the function $\boldsymbol{h}$ from data, or to develop a summary score similar to eBIC that can be used to compare not just different tuning parameters but also across different models.

\subsubsection*{References}
\bibliographystyle{abbrvnat}
\begingroup
\renewcommand{\section}[2]{}%
\bibliography{Final}
\endgroup

\clearpage


\section*{Supplementary material}

# Supplemental Information: Graphical Models for Non-Negative Data Using Generalized Score Matching

**Shiqing Yu**  
University of Washington

**Mathias Drton**  
University of Washington

**Ali Shojaie**  
University of Washington

## A.2 SCORE MATCHING

The following lemma is used in the proof of Theorem 2.

**Lemma A.1.** *Assuming that $f$ and $g$ are differentiable a.e., then for all $j = 1, \ldots, m$,*

$$\lim_{a \nearrow +\infty, b \searrow 0^+} f(\boldsymbol{x}_{-j}; a)g(\boldsymbol{x}_{-j}; a) - f(\boldsymbol{x}_{-j}; b)g(\boldsymbol{x}_{-j}; b) = \int_0^\infty f(\boldsymbol{x})\frac{\partial g(\boldsymbol{x})}{\partial x_j}\,\mathrm{d}x_j + \int_0^\infty g(\boldsymbol{x})\frac{\partial f(\boldsymbol{x})}{\partial x_j}\,\mathrm{d}x_j,$$

*where $(\boldsymbol{x}_{-j}; a)$ is the vector obtained by replacing the $j$-th component of $\boldsymbol{x}$ by $a$.*

*Proof.* This is just an analog of Lemma 4 from Hyvärinen (2005) proved by integrating the partial derivatives. □

*Proof of Theorem 2.* Recall the following assumptions given in Section 2.3.

(A1)  $p_0(\boldsymbol{x})h_j(x_j)\partial_j \log p(\boldsymbol{x}) \to 0$ as $x_j \nearrow +\infty$ and as $x_j \searrow 0^+$, $\forall \boldsymbol{x}_{-j} \in \mathbb{R}_+^{m-1}$, $\forall p \in \mathcal{P}_+$,

(A2)  $\mathbb{E}_{p_0}\|\nabla \log p(\boldsymbol{X}) \circ \boldsymbol{h}^{1/2}(\boldsymbol{X})\|_2^2 < +\infty$, $\quad \mathbb{E}_{p_0}\|(\nabla \log p(\boldsymbol{X}) \circ \boldsymbol{h}(\boldsymbol{X}))'\|_1 < +\infty$, $\quad \forall p \in \mathcal{P}_+$,

where

$$\partial_j \log p(\boldsymbol{x}) \equiv \left.\frac{\partial \log p(\boldsymbol{y})}{\partial y_j}\right|_{\boldsymbol{y}=\boldsymbol{x}}.$$

Without explicitly writing the domains $\mathbb{R}_+$ or $\mathbb{R}_+^m$ in all integrals, by (4) we have

$$J_{\boldsymbol{h}}(p) = \frac{1}{2}\int p_0(\boldsymbol{x})\left[\|\nabla \log p(\boldsymbol{x}) \circ \boldsymbol{h}^{1/2}(\boldsymbol{x})\|_2^2 - 2(\nabla \log p(\boldsymbol{x}) \circ \boldsymbol{h}^{1/2}(\boldsymbol{x}))^\top(\nabla \log p_0(\boldsymbol{x}) \circ \boldsymbol{h}^{1/2}(\boldsymbol{x}))\right.$$
$$\left. + \|\nabla \log p_0(\boldsymbol{x}) \circ \boldsymbol{h}^{1/2}(\boldsymbol{x})\|_2^2\right]\mathrm{d}\boldsymbol{x}$$

$$= \underbrace{\frac{1}{2}\int p_0(\boldsymbol{x}) \sum_{j=1}^m h_j(x_j)\left(\frac{\partial \log p(\boldsymbol{x})}{\partial x_j}\right)^2 \mathrm{d}\boldsymbol{x}}_{\equiv A} - \underbrace{\int p_0(\boldsymbol{x}) \sum_{j=1}^m h_j(x_j)\frac{\partial \log p(\boldsymbol{x})}{\partial x_j}\frac{\partial \log p_0(\boldsymbol{x})}{\partial x_j}\,\mathrm{d}\boldsymbol{x}}_{\equiv B}$$

$$+ \underbrace{\frac{1}{2}\int p_0(\boldsymbol{x}) \sum_{j=1}^m h_j(x_j)\left(\frac{\partial \log p_0(\boldsymbol{x})}{\partial x_j}\right)^2 \mathrm{d}\boldsymbol{x}}_{\equiv C},$$

where $A$ will simply appear in the final display as is, $C$ is a constant as it only involves the true pdf $p_0$, and we wish to simplify $B$ by integration by parts. We can split the integral into these three parts since $A$ and $C$ are assumed finite in the first part of (A2), and the integrand in $B$ is integrable since $|2ab| \leq a^2 + b^2$. Thus, by linearity and Fubini's theorem, we can write

1

$$B = -\sum_{j=1}^{m} \int p_0(\boldsymbol{x}) h_j(x_j) \frac{\partial \log p(\boldsymbol{x})}{\partial x_j} \frac{\partial \log p_0(\boldsymbol{x})}{\partial x_j} \,\mathrm{d}\boldsymbol{x}$$

$$= -\sum_{j=1}^{m} \int \left[ \int p_0(\boldsymbol{x}) h_j(x_j) \frac{\partial \log p(\boldsymbol{x})}{\partial x_j} \frac{\partial \log p_0(\boldsymbol{x})}{\partial x_j} \,\mathrm{d}x_j \right] \mathrm{d}\boldsymbol{x}_{-j}.$$

By the fact that $\frac{\partial \log p_0(\boldsymbol{x})}{\partial x_j} = \frac{1}{p_0(\boldsymbol{x})} \frac{\partial p_0(\boldsymbol{x})}{\partial x_j}$, this can be simplified to

$$B = -\sum_{j=1}^{m} \int \left[ \int \frac{\partial p_0(\boldsymbol{x})}{\partial x_j} h_j(x_j) \frac{\partial \log p(\boldsymbol{x})}{\partial x_j} \,\mathrm{d}x_j \right] \mathrm{d}\boldsymbol{x}_{-j}.$$

Then by Lemma A.1 and assumption (A1),

$$B = -\sum_{j=1}^{m} \int \left[ \lim_{a \nearrow \infty, b \searrow 0^+} [p_0(\boldsymbol{x}_{-j}; a) h_j(a) \partial_j \log p(\boldsymbol{x}_{-j}, a) - p_0(\boldsymbol{x}_{-j}; b) h_j(b) \partial_j \log p(\boldsymbol{x}_{-j}, b)] \right.$$

$$\left. - \int p_0(\boldsymbol{x}) \frac{\partial (h_j(x_j) \partial_j \log p(\boldsymbol{x}))}{\partial x_j} \,\mathrm{d}x_j \right] \mathrm{d}\boldsymbol{x}_{-j}$$

$$= \sum_{j=1}^{m} \int \left[ \int p_0(\boldsymbol{x}) \frac{\partial (h_j(x_j) \partial_j \log p(\boldsymbol{x}))}{\partial x_j} \,\mathrm{d}x_j \right] \mathrm{d}\boldsymbol{x}_{-j}.$$

Justified by the second half of (A2), by Fubini-Tonelli and linearity again

$$B = \sum_{j=1}^{m} \int p_0(\boldsymbol{x}) \frac{\partial (h_j(x_j) \partial_j \log p(\boldsymbol{x}))}{\partial x_j} \,\mathrm{d}\boldsymbol{x},$$

$$= \sum_{j=1}^{m} \int h_j'(x_j) \frac{\partial \log p(\boldsymbol{x})}{\partial x_j} p_0(\boldsymbol{x}) \,\mathrm{d}\boldsymbol{x} + \sum_{j=1}^{m} \int h_j(x_j) \frac{\partial^2 \log p(\boldsymbol{x})}{\partial x_j^2} p_0(\boldsymbol{x}) \,\mathrm{d}\boldsymbol{x}.$$

Thus,

$$J_{\boldsymbol{h}}(p) = B + A + C$$

$$= \int_{\mathbb{R}_+^m} p_0(\boldsymbol{x}) \sum_{j=1}^{m} \left[ h_j'(x_j) \frac{\partial \log p(\boldsymbol{x})}{\partial x_j} + h_j(x_j) \frac{\partial^2 \log p(\boldsymbol{x})}{\partial x_j^2} + \frac{1}{2} h_j(x_j) \left( \frac{\partial \log p(\boldsymbol{x})}{\partial x_j} \right)^2 \right] \mathrm{d}\boldsymbol{x} + C,$$

where $C$ is a constant that does not depend on $p$. $\square$

*Proof of Theorem 3.* By definition $J_{\boldsymbol{h}}(p_{\boldsymbol{\theta}}) \geq 0$ and $J_{\boldsymbol{h}}(p_{\boldsymbol{\theta}_0}) = 0$, so $\boldsymbol{\theta}_0$ minimizes $J_{\boldsymbol{h}}(p_{\boldsymbol{\theta}})$. Conversely, suppose $J_{\boldsymbol{h}}(p_{\boldsymbol{\theta}}) = 0$ for some $\boldsymbol{\theta}_1 \in \boldsymbol{\Theta}$. By assumption $p_0(\boldsymbol{x}) > 0$ almost surely (hereafter a.s.) and $h_j^{1/2}(\boldsymbol{x}) > 0$ a.s. for all $j = 1, \ldots, m$. Therefore, we must have $\nabla \log p_{\boldsymbol{\theta}_1}(\boldsymbol{x}) = \nabla \log p_0(\boldsymbol{x})$ a.s., or equivalently, $p_{\boldsymbol{\theta}_1}(\boldsymbol{x}) = \mathrm{const} \times p_0(\boldsymbol{x})$ for all almost every $\boldsymbol{x} \in \mathbb{R}_+^m$. Since $p_{\boldsymbol{\theta}_1}$ and $p_0$ are both continuous probability density functions, we necessarily have $p_{\boldsymbol{\theta}_1}(\boldsymbol{x}) = p_0(\boldsymbol{x})$ for all $\boldsymbol{x} \in \mathbb{R}_+^m$, which implies $\boldsymbol{\theta}_1 = \boldsymbol{\theta}_0$ by the identifiability assumption. The last claim follows by the law of large numbers, and is an analog of Corollary 3 in Hyvärinen (2005). $\square$

## A.3 EXPONENTIAL FAMILIES

Consider the case where $\{p_{\boldsymbol{\theta}} : \boldsymbol{\theta} \in \boldsymbol{\Theta} \subseteq \mathbb{R}^r\}$ contains exponential families with densities

$$\log p_{\boldsymbol{\theta}}(\boldsymbol{x}) = \boldsymbol{\theta}^\top \boldsymbol{t}(\boldsymbol{x}) - \psi(\boldsymbol{\theta}) + b(\boldsymbol{x}), \quad \boldsymbol{x} \in \mathbb{R}_+^m.$$

Then the empirical generalized $\boldsymbol{h}$-score matching loss becomes

$$\hat{J}_{\boldsymbol{h}}(p_{\boldsymbol{\theta}}) = \frac{1}{2} \boldsymbol{\theta}^\top \boldsymbol{\Gamma}(\mathbf{x}) \boldsymbol{\theta} - \boldsymbol{g}(\mathbf{x})^\top \boldsymbol{\theta} + \mathrm{const},$$

where

$$\boldsymbol{\Gamma}(\mathbf{x}) = \frac{1}{n}\sum_{i=1}^{n}\sum_{j=1}^{m} h_j(X_j^{(i)})\boldsymbol{t}'_j(\boldsymbol{X}^{(i)})\boldsymbol{t}'_j(\boldsymbol{X}^{(i)})^\top \in \mathbb{R}^{r\times r} \quad \text{and} \tag{A.1}$$

$$\boldsymbol{g}(\mathbf{x}) = -\frac{1}{n}\sum_{i=1}^{n}\left[h_j(X_j^{(i)})b'_j(\boldsymbol{X}^{(i)})\boldsymbol{t}'_j(\boldsymbol{X}^{(i)}) + h_j(X_j^{(i)})\boldsymbol{t}''_j(\boldsymbol{X}^{(i)}) + h'_j(X_j^{(i)})\boldsymbol{t}'_j(\boldsymbol{X}_i)\right] \in \mathbb{R}^r. \tag{A.2}$$

*Proof of (6).* For exponential families, under the assumptions the empirical loss $\hat{J}_{\boldsymbol{h}}(p_{\boldsymbol{\theta}})$ becomes (up to an additive constant)

$$\begin{aligned}
\hat{J}_{\boldsymbol{h}}(p_{\boldsymbol{\theta}}) &= \frac{1}{n}\sum_{i=1}^{n}\sum_{j=1}^{m}\left[h'_j(X_j^{(i)})\frac{\partial \log p_{\boldsymbol{\theta}}(\boldsymbol{X}^{(i)})}{\partial X_j^{(i)}} + h_j(X_j^{(i)})\frac{\partial^2 \log p_{\boldsymbol{\theta}}(\boldsymbol{X}^{(i)})}{\partial (X_j^{(i)})^2} + \frac{1}{2}h_j(X_j^{(i)})\left(\frac{\partial \log p_{\boldsymbol{\theta}}(\boldsymbol{X}^{(i)})}{\partial X_j^{(i)}}\right)^2\right]\\
&= \frac{1}{n}\sum_{i=1}^{n}\sum_{j=1}^{m}\left[h'_j(X_j^{(i)})(\boldsymbol{\theta}^\top \boldsymbol{t}'_j(\boldsymbol{X}^{(i)}) + b'_j(\boldsymbol{X}^{(i)})) + h_j(X_j^{(i)})(\boldsymbol{\theta}^\top \boldsymbol{t}''_j(\boldsymbol{X}^{(i)}) + b''_j(\boldsymbol{X}^{(i)}))\right.\\
&\qquad \left. + \frac{1}{2}h_j(X_j^{(i)})(\boldsymbol{\theta}^\top \boldsymbol{t}'_j(\boldsymbol{X}^{(i)}) + b'_j(\boldsymbol{X}^{(i)}))^2\right]\\
&= \frac{1}{n}\left\{\frac{1}{2}\boldsymbol{\theta}^\top\left[\sum_{i=1}^{n}\sum_{j=1}^{m}h_j(X_j^{(i)})\boldsymbol{t}'_j(\boldsymbol{X}^{(i)})\boldsymbol{t}'_j(\boldsymbol{X}^{(i)})^\top\right]\boldsymbol{\theta}\right.\\
&\qquad \left. + \left[\sum_{i=1}^{n}h_j(X_j^{(i)})b'_j(\boldsymbol{X}^{(i)})\boldsymbol{t}'_j(\boldsymbol{X}^{(i)}) + h_j(X_j^{(i)})\boldsymbol{t}''_j(\boldsymbol{X}^{(i)}) + h'_j(X_j^{(i)})\boldsymbol{t}'_j(\boldsymbol{X}^{(i)})\right]^\top \boldsymbol{\theta}\right\} + \text{const},
\end{aligned}$$

which is quadratic in $\boldsymbol{\theta}$. Let

$$\boldsymbol{\Gamma}(\mathbf{x}) = \frac{1}{n}\sum_{i=1}^{n}\sum_{j=1}^{m} h_j(X_j^{(i)})\boldsymbol{t}'_j(\boldsymbol{X}^{(i)})\boldsymbol{t}'_j(\boldsymbol{X}^{(i)})^\top, \tag{A.3}$$

$$\boldsymbol{g}(\mathbf{x}) = -\frac{1}{n}\sum_{i=1}^{n}\left[h_j(X_j^{(i)})b'_j(\boldsymbol{X}^{(i)})\boldsymbol{t}'_j(\boldsymbol{X}^{(i)}) + h_j(X_j^{(i)})\boldsymbol{t}''_j(\boldsymbol{X}^{(i)}) + h'_j(X_j^{(i)})\boldsymbol{t}'_j(\boldsymbol{X}_i)\right]. \tag{A.4}$$

Then we can write $\hat{J}_{\boldsymbol{h}}(p_{\boldsymbol{\theta}}) = \frac{1}{2}\boldsymbol{\theta}^\top \boldsymbol{\Gamma}(\mathbf{x})\boldsymbol{\theta} - \boldsymbol{g}(\mathbf{x})^\top \boldsymbol{\theta} + \text{const}$. □

*Proof of Theorem 4.* Recall that $\hat{J}_{\boldsymbol{h}}(p_{\boldsymbol{\theta}}) = \frac{1}{2}\boldsymbol{\theta}^\top \boldsymbol{\Gamma}\boldsymbol{\theta} - \boldsymbol{g}^\top \boldsymbol{\theta} + \text{const}$. The minimizer of $\hat{J}_{\boldsymbol{h}}(p_{\boldsymbol{\theta}})$ is thus available in the unique closed form $\hat{\boldsymbol{\theta}} \equiv \boldsymbol{\Gamma}(\mathbf{x})^{-1}\boldsymbol{g}(\mathbf{x})$ as long as $\boldsymbol{\Gamma}$ is invertible (C1). Since $\boldsymbol{\Gamma}$ and $\boldsymbol{g}$ are sample averages, by Khinchin's weak law of large numbers we have $\boldsymbol{\Gamma} \to_p \mathbb{E}_{p_0}\boldsymbol{\Gamma} \equiv \boldsymbol{\Gamma}_0$ and $\boldsymbol{g} \to_p \mathbb{E}_{p_0}\boldsymbol{g} \equiv \boldsymbol{g}_0$, where existence of $\boldsymbol{\Gamma}_0$ and $\boldsymbol{g}_0$ is assumed in (C2). Since $J_{\boldsymbol{h}}(p_{\boldsymbol{\theta}}) = \mathbb{E}[\hat{J}_{\boldsymbol{h}}(p_{\boldsymbol{\theta}})] = \mathbb{E}[\frac{1}{2}\boldsymbol{\theta}^\top \boldsymbol{\Gamma}(\mathbf{x})\boldsymbol{\theta} - \boldsymbol{g}(\mathbf{x})^\top \boldsymbol{\theta}] = \frac{1}{2}\boldsymbol{\theta}^\top \boldsymbol{\Gamma}_0\boldsymbol{\theta} - \boldsymbol{g}_0\boldsymbol{\theta}$ and we know $\boldsymbol{\theta}_0$ minimizes $J_{\boldsymbol{h}}(p_{\boldsymbol{\theta}})$ by definition, by first-order condition we munst have $\boldsymbol{\Gamma}_0\boldsymbol{\theta}_0 = \boldsymbol{g}_0$. Then by Lindeberg-Lévy central limit theorem (recall that $\boldsymbol{g}(\mathbf{x})$ and $\boldsymbol{\Gamma}(\mathbf{x})$ are sample averages)

$$\sqrt{n}(\boldsymbol{g}(\mathbf{x}) - \boldsymbol{\Gamma}(\mathbf{x})\boldsymbol{\theta}_0) \to_d \mathcal{N}_m(\mathbf{0}, \boldsymbol{\Sigma}_0),$$

where $\boldsymbol{\Sigma}_0 \equiv \mathbb{E}_{p_0}[(\boldsymbol{\Gamma}(\mathbf{x})\boldsymbol{\theta}_0 - \boldsymbol{g}(\mathbf{x}))(\boldsymbol{\Gamma}(\mathbf{x})\boldsymbol{\theta}_0 - \boldsymbol{g}(\mathbf{x}))^\top]$, as long as $\boldsymbol{\Sigma}_0$ exists (C2).
Then by Slutsky's theorem,

$$\sqrt{n}(\hat{\boldsymbol{\theta}} - \boldsymbol{\theta}_0) \equiv \sqrt{n}(\boldsymbol{\Gamma}(\mathbf{x})^{-1}(\boldsymbol{g}(\mathbf{x}) - \boldsymbol{\Gamma}(\mathbf{x})\boldsymbol{\theta}_0)) \to_d \mathcal{N}_r(\mathbf{0}, \boldsymbol{\Gamma}_0^{-1}\boldsymbol{\Sigma}\boldsymbol{\Gamma}_0^{-1}),$$

as long as $\boldsymbol{\Gamma}_0$ is invertible (C2).
For the second half of the theorem, (C2) $\mathbb{E}_{p_0}\boldsymbol{\Gamma}(\mathbf{x}) < \infty$ and $\mathbb{E}_{p_0}\boldsymbol{g}(\mathbf{x}) < \infty$ implies $\mathbb{E}_{p_0}|\boldsymbol{\Gamma}(\mathbf{x})| < \infty$ and $\mathbb{E}_{p_0}|\boldsymbol{g}(\mathbf{x})| < \infty$, so by strong law of large numbers (and a union bound on at most $k^2$ null sets)

$$\boldsymbol{\Gamma}(\mathbf{x}) \to_{\text{a.s.}} \boldsymbol{\Gamma}_0, \quad \boldsymbol{g}(\mathbf{x}) \to_{\text{a.s.}} \boldsymbol{g}_0.$$

Then outside a null set,
$$\hat{\boldsymbol{\theta}} \equiv \boldsymbol{\Gamma}(\mathbf{x})^{-1}\boldsymbol{g}(\mathbf{x}) \to_{\text{a.s.}} \boldsymbol{\Gamma}_0^{-1}\boldsymbol{g}_0 = \boldsymbol{\theta}_0.$$

□

*Proof for Example 5.* For the family of univariate truncated Gaussian distributions with unknown mean parameter $\mu$ and known variance parameter $\sigma^2$, we have

$$p_\theta(x) \propto \exp\left(\theta t(x) + b(x)\right), \quad \theta \equiv \frac{\mu}{\sigma^2}, \quad t(x) \equiv x, \quad b(x) = -\frac{x^2}{2\sigma^2}.$$

We choose to estimate $\theta \equiv \mu/\sigma^2$. Then by (A.1) and (A.2),

$$\hat{\mu}_h = \sigma^2 \hat{\theta} \equiv \sigma^2 \Gamma(\boldsymbol{x})^{-1} g(\boldsymbol{x})$$
$$= -\sigma^2 \left[\sum_{i=1}^n h(X_i)t'(X_i)^2\right]^{-1} \left[\sum_{i=1}^n h(X_i)b'(X_i)t'(X_i) + h(X_i)t''(X_i) + h'(X_i)t'(X_i)\right]$$
$$= -\sigma^2 \left[\sum_{i=1}^n h(X_i)\right]^{-1} \left[\sum_{i=1}^n -h(X_i)\frac{X_i}{\sigma^2} + h'(X_i)\right].$$

By Theorem 4,

$$\sqrt{n}(\hat{\mu}_h - \mu_0) \to_d \mathcal{N}\left(0, \frac{\sigma^4 \mathbb{E}_0\left[h(X)\frac{\mu_0-X}{\sigma^2} + h'(X)\right]^2}{\mathbb{E}_0^2[h(X)]}\right) \sim \mathcal{N}\left(0, \frac{\mathbb{E}_0\left[h(X)(\mu_0-X) + \sigma^2 h'(X)\right]^2}{\mathbb{E}_0^2[h(X)]}\right).$$

By integration by parts, (suppressing the dependence of $p_{\mu_0}$ on $\mu_0$)

$$\mathbb{E}_0[h(X)h'(X)(X-\mu_0)]$$
$$= \int_0^\infty h'(x)h(x)(x-\mu_0)p(x)\,dx = \int_0^\infty h(x)(x-\mu_0)p(x)\,dh(x)$$
$$= h^2(x)(x-\mu_0)p(x)\big|_0^\infty - \int h(x)\,dh(x)(x-\mu_0)p(x)$$
$$= -\int h^2(x)p(x)\,dx - \int h(x)h'(x)(x-\mu_0)p(x)\,dx + \int h^2(x)\frac{(x-\mu_0)^2}{\sigma^2}p(x)\,dx,$$

where the last step follows from the assumptions $\lim_{x\searrow 0^+} h(x) = 0$ and $\lim_{x\nearrow +\infty} h^2(x)(x-\mu_0)p_{\mu_0}(x) = 0$. So

$$\mathbb{E}_0[h(X)h'(X)(X-\mu_0)] = \frac{\mathbb{E}[h^2(X)((X-\mu_0)^2/\sigma^2 - 1)]}{2}. \tag{A.5}$$

The asymptotic variance thus becomes

$$\frac{\mathbb{E}_0\left[h(X)(\mu_0-X) + \sigma^2 h'(X)\right]^2}{\mathbb{E}_0^2[h(X)]}$$
$$= \frac{\mathbb{E}_0\left[h^2(X)(X-\mu_0)^2 - 2\sigma^2 h^2(X)\left((X-\mu_0)^2/\sigma^2 - 1\right)/2 + \sigma^4 h'^2(X)\right]}{\mathbb{E}_0^2[h(X)]}$$
$$= \frac{\mathbb{E}_0[\sigma^2 h^2(X) + \sigma^4 h'^2(X)]}{\mathbb{E}_0^2[h(X)]}.$$

We note that the Cramér-Rao lower bound is $\frac{\sigma^4}{\text{var}(X-\mu_0)}$, which follows from taking the second derivative of $\log p_{\mu_0}$ with respect to $\mu_0$.

□

*Proof for Example 6.* For the family of univariate truncated Gaussian distributions with known mean parameter $\mu$ and unknown variance parameter $\sigma^2 > 0$, we have

$$p_\theta(x) \propto \exp\left(\theta t(x) + b(x)\right), \quad \theta \equiv \frac{1}{\sigma^2}, \quad t(x) \equiv -(x-\mu)^2/2, \quad b(x) = 0.$$

We estimate $\theta \equiv 1/\sigma^2$. By (A.1) and (A.2),

$$\hat\theta \equiv \Gamma(\boldsymbol{x})^{-1}g(\boldsymbol{x})$$

$$= -\left[\sum_{i=1}^n h(X_i)t'(X_i)^2\right]^{-1}\left[\sum_{i=1}^n h(X_i)b'(X_i)t'(X_i) + h(X_i)t''(X_i) + h'(X_i)t'(X_i)\right]$$

$$= \left[\sum_{i=1}^n h(X_i)(X_i-\mu)^2\right]^{-1}\left[\sum_{i=1}^n h(X_i) + h'(X_i)(X_i-\mu)\right].$$

By Theorem 4, $\sqrt{n}(\hat\theta - \theta) \to_d \mathcal{N}(0, \varsigma^2)$, where

$$\varsigma^2 \equiv \frac{\mathbb{E}_0\left[h(X)((X-\mu)^2/\sigma_0^2 - 1) - h'(X)(X-\mu)\right]^2}{\mathbb{E}_0^2[h(X)(X-\mu)^2]}$$

$$= \frac{1}{\mathbb{E}_0^2[h(X)(X-\mu)^2]}\bigg(\mathbb{E}_0[h^2(X)(X-\mu)^4/\sigma_0^4 - 2h^2(X)(X-\mu)^2/\sigma_0^2 + h^2(X) + h'^2(X)(X-\mu)^2$$

$$- 2h(X)h'(X)(X-\mu)^3/\sigma_0^2 + 2h(X)h'(X)(X-\mu)\bigg).$$

By integration by parts, (suppressing the dependence of $p_{\sigma_0^2}$ on $\sigma_0^2$)

$$\mathbb{E}_0[h(X)h'(X)(X-\mu)^3]$$

$$= \int_0^\infty h'(x)h(x)(x-\mu)^3 p(x)\,\mathrm{d}x = \int_0^\infty h(x)(x-\mu)^3 p(x)\,\mathrm{d}h(x)$$

$$= h^2(x)(x-\mu)^3 p(x)\big|_0^\infty - \int h(x)\,\mathrm{d}h(x)(x-\mu)^3 p(x)$$

$$= -\int h(x)h'(x)(x-\mu)^3 p(x)\,\mathrm{d}x - 3\int h^2(x)(x-\mu)^2 p(x)\,\mathrm{d}x + \int h^2(x)\frac{(x-\mu)^4}{\sigma_0^2}p(x)\,\mathrm{d}x,$$

where the last step follows from the assumptions $\lim_{x \searrow 0^+} h(x) = 0$ and $\lim_{x \nearrow +\infty} h^2(x)(x-\mu)^3 p_{\sigma_0^2}(x) = 0$. Combining this with (A.5) we get

$$\sqrt{n}(\hat\theta - \theta) \to_d \mathcal{N}(0, \varsigma^2) \sim \mathcal{N}\left(0, \frac{2\mathbb{E}_0[h^2(X)(X-\mu)^2/\sigma_0^2] + \mathbb{E}_0[h'^2(X-\mu)^2]}{\mathbb{E}_0^2[h(X)(X-\mu)^2]}\right),$$

and so by the delta method, for $\hat\sigma_k^2 \equiv \hat\theta^{-1}$,

$$\sqrt{n}(\hat\sigma_h^2 - \sigma_0^2) \to_d \mathcal{N}\left(0, \frac{2\sigma_0^6 \mathbb{E}_0[h^2(X)(X-\mu)^2] + \sigma_0^8 \mathbb{E}_0[h'^2(X-\mu)^2]}{\mathbb{E}_0^2[h(X)(X-\mu)^2]}\right).$$

We note that the Cramér-Rao lower bound is $\frac{4\sigma_0^8}{\mathrm{var}(X-\mu)^2}$, which follows from taking the second derivative of $\log p_{\sigma_0^2}$ with respect to $\sigma_0^2$. $\square$

## A.4 REGULARIZED GENERALIZED SCORE MATCHING

We first verify assumptions (A1)–(A2) in the case of truncated Gaussian distributions.

**Lemma A.2** (Assumptions for truncated Gaussian). *Consider the non-centered truncated Gaussian distribution with density*
$$\log p_0(\boldsymbol{x}) = -\frac{1}{2}(\boldsymbol{x}-\boldsymbol{\mu}_0)^\top \mathbf{K}_0(\boldsymbol{x}-\boldsymbol{\mu}_0) + \text{const}$$
*with unknown positive definite inverse covariance parameter $\mathbf{K}_0$ and unknown mean parameter $\boldsymbol{\mu}_0$. Then assuming $0 \leq h_j \leq M_j$, $\lim_{x_j \searrow 0^+} h_j(x_j) = 0$ and $|h_j'| \leq M_j'$, assumptions (A1)–(A2) for score matching are satisfied for any proposed parameters $\mathbf{K} \succ \mathbf{0}$ and $\boldsymbol{\mu}$. Taking $\boldsymbol{\mu} \equiv \boldsymbol{\mu}_0 \equiv \mathbf{0}$ the assumptions also hold in the centered setting. Choosing $m=1$ gives the univariate case.*

*Proof of Lemma A.2.* Consider $p \sim \text{TN}(\boldsymbol{\mu}, \mathbf{K})$, with $\boldsymbol{k}_j$ the $j$-th column of $\mathbf{K}$. Let $M \equiv \max_j M_j$ and $M' \equiv \max_j M_j'$.

(A1) For any fixed $\boldsymbol{x}_{-j} \in \mathbb{R}_+^{m-1}$ and any $p \in \mathcal{P}_+$ with parameters $\mathbf{K}$ and $\boldsymbol{\mu}$,

$$\lim_{x_j \nearrow \infty} h_j(x_j) p_0(\boldsymbol{x}) \partial_j \log p(\boldsymbol{x}) \propto \lim_{x_j \nearrow \infty} h_j(x_j) \exp\left(-\frac{1}{2}(\boldsymbol{x}-\boldsymbol{\mu}_0)^\top \mathbf{K}_0(\boldsymbol{x}-\boldsymbol{\mu}_0)\right) \boldsymbol{k}_j^\top(\boldsymbol{x}-\boldsymbol{\mu})$$
$$= \lim_{x_j \nearrow \infty} h_j(x_j) \exp\left(C_1 + C_2 x_j - \frac{1}{2}\kappa_{0,jj} x_j^2\right)(C_3 + C_4 x_j)$$

for some constants $C_1, C_2, C_3$, and $C_4$ depending on $\boldsymbol{x}_{-j}, \mathbf{K}_0, \mathbf{K}, \boldsymbol{\mu}_0$ and $\boldsymbol{\mu}$. Since $\kappa_{0,jj} > 0$ and we assumed $h_j$ to be bounded, the limit equals to 0 for all $j$ and $\boldsymbol{x}_{-j}$.
Similarly,

$$\lim_{x_j \searrow 0^+} h_j(x_j) p_0(\boldsymbol{x}) \partial_j \log p(\boldsymbol{x}) \propto \lim_{x_j \searrow 0^+} h_j(x_j) \exp\left(C_1 + C_2 x_j - \frac{1}{2}\kappa_{0,jj} x_j^2\right)(C_3 + C_4 x_j)$$
$$= \exp(C_1) C_3 \lim_{x_j \searrow 0^+} h_j(x_j) = 0$$

if and only if we assume $\lim_{x_j \searrow 0^+} h_j(x_j) = 0$.

(A2) For any $p \in \mathcal{P}_+$ with parameters $\mathbf{K}$ and $\boldsymbol{\mu}$,

$$\mathbb{E}_{p_0} \|\nabla \log p(\boldsymbol{X}) \circ \boldsymbol{h}^{1/2}(\boldsymbol{X})\|_2^2 \leq M \mathbb{E}_{p_0} \|\nabla \log p(\boldsymbol{X})\|_2^2 = M \text{tr}\left(\mathbb{E}_{p_0}\left[(\mathbf{K}(\boldsymbol{X}-\boldsymbol{\mu}))(\mathbf{K}(\boldsymbol{X}-\boldsymbol{\mu}))^\top\right]\right)$$
$$= M \text{tr}\left(\mathbf{K} \mathbb{E}_{p_0}\left[(\boldsymbol{X}-\boldsymbol{\mu}_0+(\boldsymbol{\mu}_0-\boldsymbol{\mu}))(\boldsymbol{X}-\boldsymbol{\mu}_0+(\boldsymbol{\mu}_0-\boldsymbol{\mu}))^\top\right]\mathbf{K}^\top\right)$$
$$= M \text{tr}\left(\mathbf{K}\left(\mathbf{K}_0^{-1} + (\boldsymbol{\mu}_0-\boldsymbol{\mu})(\boldsymbol{\mu}_0-\boldsymbol{\mu})^\top\right)\mathbf{K}\right) < +\infty$$

since $M, \mathbf{K}, \mathbf{K}_0, \boldsymbol{\mu}, \boldsymbol{\mu}_0$ are all finite constants. We also have

$$\mathbb{E}_{p_0} \|(\nabla \log p(\boldsymbol{X}) \circ \boldsymbol{h}(\boldsymbol{X}))'\|_1 = \sum_{i=1}^m \mathbb{E}_{p_0} \left|h_j'(X_j) \partial_j \log p(\boldsymbol{X}) + h_j(X_j) \partial_j^2 \log p(\boldsymbol{X})\right|$$
$$\leq \sum_{i=1}^m \mathbb{E}_{p_0} |h_j'(X_j) \partial_j \log p(\boldsymbol{X})| + \mathbb{E}_{p_0}|h_j(X_j) \partial_j^2 \log p(\boldsymbol{X})|$$
$$\leq \sum_{i=1}^m M' \mathbb{E}_{p_0}|\boldsymbol{k}_j^\top(\boldsymbol{X}-\boldsymbol{\mu})| + M\kappa_{jj}$$
$$\leq \sum_{i=1}^m M'|\boldsymbol{k}_j|^\top \mathbb{E}_{p_0} \boldsymbol{X} + M'|\boldsymbol{k}_j^\top \boldsymbol{\mu}| + M\text{tr}(\mathbf{K}) < +\infty.$$

Hence, (A1) and (A2) are both satisfied.

□

Our analysis of the regularized generalized $\boldsymbol{h}$-score matching estimator follows the proof for the following theorem from Lin et al. (2016), restated below. In our definition and implementation we choose to optimize over all symmetric matrices, but we adopt the following theorem in whose proof the symmetry condition is not explicitly imposed, in order to decouple the columns of $\mathbf{K}$ and to highlight the scaling.

**Theorem A.3** (Analog of Theorem 1 from Lin et al. (2016))**.** *Recall that $S_0 \equiv S(\mathbf{K}_0) \equiv \{(i,j) : \kappa_{0,ij} \neq 0\}$. Suppose $\mathbf{\Gamma}_{0,S_0 S_0}$ is invertible and satisfies the irrepresentability condition (10) with incoherence parameter $\alpha \in (0,1]$. Assume that*

$$\|\mathbf{\Gamma}(\mathbf{x}) - \mathbf{\Gamma}_0\|_\infty < \epsilon_1, \qquad \|\mathbf{g}(\mathbf{x}) - \mathbf{g}_0\|_\infty < \epsilon_2, \qquad (A.6)$$

*with $d_{\mathbf{K}_0} \epsilon_1 \leq \alpha/(6 c_{\mathbf{\Gamma}_0})$. If*

$$\lambda > \frac{3(2-\alpha)}{\alpha} \max\{c_{\mathbf{K}_0} \epsilon_1, \epsilon_2\},$$

*then the following statements hold:*

(a) *The regularized generalized $\mathbf{h}$-score matching estimator $\hat{\mathbf{K}}$ in (9) is unique, with support $\hat{S} \equiv S(\hat{\mathbf{K}}) \subseteq S_0$, and satisfies*

$$\|\hat{\mathbf{K}} - \mathbf{K}_0\|_\infty \leq \frac{c_{\mathbf{\Gamma}_0}}{2-\alpha} \lambda.$$

(b) *If*

$$\min_{1 \leq j < k \leq m} |\mathbf{K}_{0,jk}| > \frac{c_{\mathbf{\Gamma}_0}}{2-\alpha} \lambda,$$

*then $\hat{S} = S_0$ and $\text{sign}(\hat{\mathbf{K}}_{jk}) = \text{sign}(\mathbf{K}_{0.jk})$ for all $(j,k) \in S_0$.*

This is a deterministic result, and the improvement of our generalized estimator over the one in Lin et al. (2016) is in its asymptotic guarantees, as in Theorem 10. We present a corollary to this theorem, as seen in the second and third inequalities in Theorem 10 (a).

**Corollary A.1.** *Suppose the same assumptions under Theorem A.3 hold. Then $\hat{\mathbf{K}}$ satisfies*

$$\|\hat{\mathbf{K}} - \mathbf{K}_0\|_F \leq \frac{c_{\mathbf{\Gamma}_0}}{2-\alpha} \lambda \sqrt{|S_0|} \leq \frac{c_{\mathbf{\Gamma}_0}}{2-\alpha} \lambda \sqrt{d_{\mathbf{K}_0} m},$$

$$\|\hat{\mathbf{K}} - \mathbf{K}_0\|_2 \leq \frac{c_{\mathbf{\Gamma}_0}}{2-\alpha} \lambda \min(\sqrt{|S_0|}, d_{\mathbf{K}_0}).$$

*Proof of Corollary A.1.* By Theorem A.3, under assumptions in that theorem, the support of $\hat{\mathbf{K}}$ is a subset of the true support of $\mathbf{K}_0$, and $\|\hat{\mathbf{K}} - \mathbf{K}_0\|_\infty \leq \frac{c_{\mathbf{\Gamma}_0}}{2-\alpha} \lambda$. Since $\mathbf{K}_0$ has $|S_0|$ nonzero entries,

$$\|\hat{\mathbf{K}} - \mathbf{K}_0\|_F = \left[\sum_{\mathbf{K}_{0,jk} \neq 0} (\hat{\mathbf{K}}_{jk} - \mathbf{K}_{0,jk})^2\right]^{1/2} \leq \sqrt{|S_0|} \|\hat{\mathbf{K}} - \mathbf{K}_0\|_\infty \leq \frac{c_{\mathbf{\Gamma}_0}}{2-\alpha} \lambda \sqrt{|S_0|}.$$

Similarly, by the definition of matrix $\ell_\infty$-$\ell_\infty$ norm,

$$\|\|\hat{\mathbf{K}} - \mathbf{K}_0\|\|_2 \leq \|\|\hat{\mathbf{K}} - \mathbf{K}_0\|\|_\infty = \max_{j=1,\ldots,m} \sum_{k=1}^m |\hat{\mathbf{K}}_{jk} - \mathbf{K}_{0,jk}| \leq \frac{c_{\mathbf{\Gamma}_0}}{2-\alpha} \lambda d_{\mathbf{K}_0}.$$

The result follows by also noting that $\|\|\hat{\mathbf{K}} - \mathbf{K}_0\|\|_2 \leq \|\hat{\mathbf{K}} - \mathbf{K}_0\|_F$. □

*Proof of Theorem 10.* By Theorem A.3 it suffices to prove that for any $\tau > 3$, we can bound $\|\mathbf{\Gamma}(\mathbf{x}) - \mathbf{\Gamma}_0\|_\infty$ by some $\epsilon_1$ and $\|\mathbf{g}(\mathbf{x}) - \mathbf{g}_0\|_\infty$ by some $\epsilon_2$, uniformly with probability $1 - m^{3-\tau}$. Recall from Section 4.2 that the $j^{\text{th}}$ block of $\mathbf{\Gamma} \in \mathbb{R}^{m^2 \times m^2}$ has $(k, \ell)$-th entry

$$\frac{1}{n} \sum_{i=1}^n X_k^{(i)} X_\ell^{(i)} h_j(X_j^{(i)}),$$

and the entry in $\mathbf{g} \in \mathbb{R}^{m^2}$ (obtained by linearizing a $m \times m$ matrix) corresponding to $(j, k)$ with $j \neq k$, is

$$\frac{1}{n} \sum_{i=1}^n X_k^{(i)} h_j'(X_j^{(i)}),$$

while the entry for $(j,j)$ is

$$\frac{1}{n}\sum_{i=1}^n X_j^{(i)} h_j'(X_j^{(i)}) + \frac{1}{n}\sum_{i=1}^n h_j(X_j^{(i)}).$$

Denote $M \equiv \max_j \sup_{x>0} h_j(x)$ and $M' \equiv \max_j \sup_{x>0} h_j'(x)$, and let $c_{\mathbf{X}} \equiv 2\max_j(2\sqrt{\Sigma_{jj}} + \sqrt{e}\mathbb{E}_0 X_j)$. Using results for sub-gaussian random variables from Lemma A.6 below and Hoeffding's inequality, we have for any $t_1, t_{2,1}, t_{2,2} > 0$,

$$\mathbb{P}\left(\left|\frac{1}{n}\sum_{i=1}^n X_k^{(i)} X_\ell^{(i)} h_j(X_j^{(i)}) - \mathbb{E}_0 X_k X_\ell h_j(X_j)\right| > t_1\right) \leq 2\exp\left(-\min\left(\frac{nt_1^2}{2M^2 c_{\mathbf{X}}^4}, \frac{nt_1}{2Mc_{\mathbf{X}}^2}\right)\right),$$

$$\mathbb{P}\left(\left|\frac{1}{n}\sum_{i=1}^n X_k^{(i)} h_j'(X_j^{(i)}) - \mathbb{E}_0 X_k h_j'(X_j)\right| \geq t_{2,1}\right) \leq 2\exp\left(-\frac{nt_{2,1}^2}{2M'^2 c_{\mathbf{X}}^2}\right),$$

$$\mathbb{P}\left(\left|\frac{1}{n}\sum_{i=1}^n h_j(X_j^{(i)}) - \mathbb{E}_0 h_j(X_j)\right| \geq t_{2,2}\right) \leq 2\exp\left(-2nt_{2,2}^2/M^2\right).$$

Choosing

$$\epsilon_1 \equiv Mc_{\mathbf{X}}^2 \max\left\{\frac{2(\log m^\tau + \log 6)}{n}, \sqrt{\frac{2(\log m^\tau + \log 6)}{n}}\right\},$$

$$\epsilon_{2,1} \equiv \sqrt{2}M'c_{\mathbf{X}}\sqrt{\frac{\log m^{\tau-1} + \log 6}{n}}, \quad \epsilon_{2,2} \equiv M\sqrt{\frac{\log m^{\tau-2} + \log 6}{2n}},$$

and taking union bounds over $m^3$, $m^2$, and $m$ events, respectively, we have

$$\mathbb{P}\left(\sup_{j,k,\ell}\left|\frac{1}{n}\sum_{i=1}^n X_k^{(i)} X_\ell^{(i)} h_j(X_j^{(i)}) - \mathbb{E}_0 X_k X_\ell h_j(X_j)\right| \geq \epsilon_1\right) \leq \frac{1}{3m^{\tau-3}},$$

$$\mathbb{P}\left(\sup_{j,k}\left|\frac{1}{n}\sum_{i=1}^n X_k^{(i)} h_j'(X_j^{(i)}) - \mathbb{E}_0 X_k h_j'(X_j)\right| \geq \epsilon_{2,1}\right) \leq \frac{1}{3m^{\tau-3}},$$

$$\mathbb{P}\left(\sup_j\left|\frac{1}{n}\sum_{i=1}^n h_j(X_j^{(i)}) - \mathbb{E}_0 h_j(X_j)\right| \geq \epsilon_{2,2}\right) \leq \frac{1}{3m^{\tau-3}}.$$

Hence, with probability at least $1 - m^{3-\tau}$, $\|\mathbf{\Gamma}(\mathbf{x}) - \mathbf{\Gamma}_0\|_\infty < \epsilon_1$ and $\|\mathbf{g}(\mathbf{x}) - \mathbf{g}_0\|_\infty < \epsilon_2 \equiv \epsilon_{2,1} + \epsilon_{2,2}$. Consider any $\tau > 3$, and let

$$c_2 \equiv \frac{6}{\alpha}c_{\mathbf{\Gamma}_0}, \qquad n \geq \max\{2M^2 c_{\mathbf{X}}^4 c_2^2 d_{\mathbf{K}_0}^2(\tau\log m + \log 6), 2Mc_{\mathbf{X}}^2 c_2 d_{\mathbf{K}_0}(\tau\log m + \log 6)\},$$

$$\lambda > \frac{3(2-\alpha)}{\alpha}\max\{c_{\mathbf{K}_0}\epsilon_1, \epsilon_2\}$$

$$\equiv \frac{3(2-\alpha)}{\alpha}\max\left\{Mc_{\mathbf{K}_0}c_{\mathbf{X}}^2\frac{2(\log m^\tau + \log 6)}{n},\right.$$

$$\left. Mc_{\mathbf{K}_0}c_{\mathbf{X}}^2\sqrt{\frac{2(\log m^\tau + \log 6)}{n}}, \sqrt{2}M'c_{\mathbf{X}}\sqrt{\frac{\log m^{\tau-1} + \log 6}{n}} + M\sqrt{\frac{\log m^{\tau-2} + \log 6}{2n}}\right\}.$$

Then $d_{\mathbf{K}_0}\epsilon_1 \leq \alpha/(6c_{\mathbf{\Gamma}_0})$ and the results follow from Theorem A.3. □

We now present the definition of sub-Gaussian and sub-exponential norms and variables as well as lemmas required for the proof above.

**Definition A.4** (Sub-Gaussian and Sub-Exponential Variables). The *sub-gaussian* $(r=2)$ and *sub-exponential* $(r=1)$ *norms* of a random variable are defined as

$$\|X\|_{\psi_r} \equiv \sup_{q\geq 1} q^{-1/r}(\mathbb{E}|X|^{rq})^{1/(rq)} \equiv \sup_{q\geq 1} q^{-1/r}\|X\|_{rq}.$$

If $\|X\|_{\psi_2} < \infty$ we say $X$ is *sub-gaussian*; if $\|X\|_{\psi_1} < \infty$ we call $X$ *sub-exponential*.
For a *zero-mean* sub-gaussian random variable $X$ also define the *sub-gaussian parameter*

$$\tau(X) = \inf\{\tau \geq 0 : \mathbb{E}\exp(tX) \leq \exp(\tau^2 t^2/2), \forall t \in \mathbb{R}\}.$$

Note that the definition of sub-gaussian norm here allows the variable to be non-centered, and is different from the one in Vershynin (2010), which uses $\|X\|_q$ in the definition. Instead, it coincides with $\theta_2$ in Buldygin and Kozachenko (2000). The definition of the sub-gaussian parameter is the same as in Buldygin and Kozachenko (2000), and the definition of the sub-exponential norm is as in Vershynin (2010).

**Lemma A.5** (Properties of Sub-Gaussian and Sub-Exponential Variables). *Then*

1) *For any $X$ and $r = 1, 2$, $\|X - \mathbb{E}X\|_{\psi_r} \leq 2\|X\|_{\psi_r}$ and $\|X\|_{\psi_r} \leq \|X - \mathbb{E}X\|_{\psi_r} + |\mathbb{E}X|$, as long as the expectation and norms are finite.*

2) *(Buldygin and Kozachenko, 2000) $\tau(X)$ is a norm on the space of all zero-mean sub-gaussian variables; in particular, $\tau(X + Y) \leq \tau(X) + \tau(Y)$ as long as the quantities are defined and finite.*
   *If $X$ is zero-mean sub-gaussian, then $\mathrm{var}(X) \leq \tau^2(X)$, $\|X\|_{\psi_2} \leq 2\tau(X)/\sqrt{e}$, $\tau(X) \leq \sqrt{e}\|X\|_{\psi_2}$.*
   *If $X_1, \ldots, X_n$ are i.i.d. zero-mean sub-gaussian, $\tau\left(\frac{1}{n}\sum_{i=1}^n X_i\right) \leq \frac{1}{\sqrt{n}}\tau(X_i)$.*

3) *If random variables $X_1$ and $X_2$ (not necessarily independent) are sub-gaussian with $\|X_1\|_{\psi_2} \leq K_1$ and $\|X_2\|_{\psi_2} \leq K_2$, then $X_1 X_2$ is sub-exponential with $\|X_1 X_2\|_{\psi_1} \leq K_1 K_2$.*

4) *(Buldygin and Kozachenko, 2000) If $X$ is zero-mean sub-gaussian,*

$$\mathbb{E}|X|^q \leq 2(q/e)^{q/2}\tau^q(X)$$

   *for any $q > 0$.*

5) *(Buldygin and Kozachenko, 2000) If $X_1, \ldots, X_n$ are independent zero-mean sub-gaussian variables, then for any $\epsilon > 0$,*

$$\mathbb{P}(|X_1| \geq \epsilon) \leq 2\exp\left(-\frac{\epsilon^2}{2\tau^2(X_1)}\right),$$

$$\mathbb{P}\left(\left|\frac{1}{n}\sum_{i=1}^n X_i\right| > \epsilon\right) \leq 2\exp\left(-\frac{n\epsilon^2}{2\max_i \tau^2(X_i)}\right).$$

6) *(Vershynin, 2010) If $X_1, \ldots, X_n$ are independent zero-mean sub-exponential random variables with $K \geq \max_i \|X_i\|_{\psi_1}$, then for any $\epsilon > 0$,*

$$\mathbb{P}(|X_1| \geq \epsilon) \leq 2\exp\left(-\min\left(\frac{\epsilon^2}{8e^2 K^2}, \frac{\epsilon}{4eK}\right)\right),$$

$$\mathbb{P}\left(\left|\frac{1}{n}\sum_{i=1}^n X_i\right| \geq \epsilon\right) \leq 2\exp\left(-\min\left(\frac{n\epsilon^2}{8e^2 K^2}, \frac{n\epsilon}{4eK}\right)\right).$$

*Proof.* 1) For $r = 1, 2$, by triangle inequality, $\|X - \mathbb{E}X\|_{\psi_r} \leq \|X\|_{\psi_r} + \|\mathbb{E}X\|_{\psi_r} = \|X\|_{\psi_r} + |\mathbb{E}X| \leq \|X\|_{\psi_r} + \mathbb{E}|X| \leq 2\|X\|_{\psi_r}$, where in the last step we used the definition of $\|\cdot\|_{\psi_r}$ with $q = 1$ for $r = 1$ and $\mathbb{E}|X| \leq (\mathbb{E}|X|^2)^{1/2}$ with $q = 2$ for $r = 2$. On the other hand, $\|X\|_{\psi_r} \leq \|X - \mathbb{E}X\|_{\psi_r} + \|\mathbb{E}X\|_{\psi_r} = \|X - \mathbb{E}X\|_{\psi_r} + |\mathbb{E}X|$.

2) These follow from Theorems 1.2 and 1.3 and Lemmas 1.2 and 1.7 from Buldygin and Kozachenko (2000), and $\sqrt[4]{3.1}e^{9/16}/\sqrt{2} \approx 1.6467 \leq 1.6487 \approx \sqrt{e}$.

3) By Hölder's inequality (or Cauchy-Schwarz),

$$\begin{aligned}
\|X_1 X_2\|_{\psi_1} &= \sup_{q \geq 1} q^{-1}(\mathbb{E}|X_1 X_2|^q)^{1/q} = \sup_{q \geq 1} q^{-1}(\mathbb{E}|X_1^q X_2^q|)^{1/q} \\
&\leq \sup_{q \geq 1} q^{-1}\left[(\mathbb{E}|X_1|^{2q})^{1/2}(\mathbb{E}|X_2|^{2q})^{1/2}\right]^{1/q} \\
&\leq \sup_{q \geq 1}\left[q^{-1/2}(\mathbb{E}|X_1|^{2q})^{1/2q}\right]\sup_{q \geq 1}\left[q^{-1/2}(\mathbb{E}|X_2|^{2q})^{1/2q}\right] \\
&= \|X_1\|_{\psi_2}\|X_2\|_{\psi_2} \leq K_1 K_2.
\end{aligned}$$

4) This is Lemma 1.4 from Buldygin and Kozachenko (2000).

5) This is Theorem 1.5 from Buldygin and Kozachenko (2000).

6) This follows from Corollary 5.17 from Vershynin (2010). □

**Lemma A.6.** *Suppose $\boldsymbol{X}$ follows a truncated normal distribution on $\mathbb{R}_+^m$ with parameters $\boldsymbol{\mu}$ and $\boldsymbol{\Sigma} = \mathbf{K}^{-1} \succ \mathbf{0}$. Let $\boldsymbol{X}^{(1)}, \ldots, \boldsymbol{X}^{(n)}$ be i.i.d. copies of $\boldsymbol{X}$, with $j$-th component of the $i$-th copy being $X_j^{(i)}$. Then*

1. *For $j = 1, \ldots, p$, $\tau(X_j - \mathbb{E}X_j) \leq \sqrt{\Sigma_{jj}}$. That is, the sub-gaussian parameter of any marginal distribution of $\boldsymbol{X}$, after centering, is bounded by the square root of its corresponding diagonal entry in the covariance parameter $\boldsymbol{\Sigma}$. Then for any $\epsilon > 0$,*

$$\mathbb{P}\left(\left|\frac{1}{n}\sum_{i=1}^n X_j^{(i)} - \mathbb{E}X_j\right| > \epsilon\right) \leq 2\exp\left(-\frac{n\epsilon^2}{2\Sigma_{jj}}\right).$$

*In particular, if $h_0$ is a function bounded by $M_0$, then for any $\epsilon > 0$,*

$$\mathbb{P}\left(\left|\frac{1}{n}\sum_{i=1}^n X_j^{(i)} h_0(X_k^{(i)}) - \mathbb{E}X_j h_0(X_k)\right| \geq \epsilon\right) \leq 2\exp\left(-\frac{n\epsilon^2}{8M_0^2(2\sqrt{\Sigma_{jj}} + \sqrt{e}\mathbb{E}X_j)^2}\right),$$

$$\tau\left(\frac{1}{n}\sum_{i=1}^n X_j^{(i)} h_0(X_k^{(i)}) - \mathbb{E}X_j h_0(X_k)\right) \leq \frac{2M_0}{\sqrt{n}}(2\sqrt{\Sigma_{jj}} + \sqrt{e}\mathbb{E}X_j),$$

$$\left\|\frac{1}{n}\sum_{i=1}^n X_j^{(i)} h_0(X_k^{(i)}) - \mathbb{E}X_j h_0(X_k)\right\|_{\psi_2} \leq \frac{4M_0}{\sqrt{en}}(2\sqrt{\Sigma_{jj}} + \sqrt{e}\mathbb{E}X_j).$$

2. *For $j, k, \ell \in \{1, \ldots, p\}$, if $h_0$ is a function bounded by $M_0$, then with $c_{\boldsymbol{X}} \equiv 2\max_j(2\sqrt{\Sigma_{jj}} + \sqrt{e}\mathbb{E}X_j)$,*

$$\|X_j X_k h_0(X_\ell) - \mathbb{E}X_j X_k h_0(X_\ell)\|_{\psi_1} \leq \frac{M_0}{2e}c_{\boldsymbol{X}}^2. \tag{A.7}$$

*In particular, for any $\epsilon > 0$,*

$$\mathbb{P}\left(\left|\frac{1}{n}\sum_{i=1}^n X_j^{(i)} X_k^{(i)} h_0(X_\ell^{(i)}) - \mathbb{E}X_j X_k h_0(X_\ell)\right| > \epsilon\right) \leq 2\exp\left(-\min\left(\frac{n\epsilon^2}{2M_0^2 c_{\boldsymbol{X}}^4}, \frac{n\epsilon}{2M_0 c_{\boldsymbol{X}}^2}\right)\right).$$

*Proof of Lemma A.6.* 1. Without loss of generality choose $j = 1$. By the definition of sub-gaussian parameters, we need to show that for all $t \in \mathbb{R}$,

$$\mathbb{E}\exp(tX_1) \leq \exp(t^2 \Sigma_{11}/2 + t\mathbb{E}X_1),$$

which is equivalent to

$$t^2 \Sigma_{11}/2 + t\mathbb{E}X_1 - \log \mathbb{E}\exp(tX_1) \geq 0 \quad \forall t \in \mathbb{R}. \tag{A.8}$$

Since the left-hand side of (A.8) equals 0 at $t = 0$, it suffices to show that its derivative

$$t\Sigma_{11} + \mathbb{E}X_1 - \frac{\mathrm{d}\log \mathbb{E}\exp(tX_1)}{\mathrm{d}t} = t\Sigma_{11} + \mathbb{E}X_1 - \frac{\frac{\mathrm{d}\mathbb{E}\exp(tX_1)}{\mathrm{d}t}}{\mathbb{E}\exp(tX_1)} \tag{A.9}$$

is non-negative on $(0, \infty)$ and non-positive on $(-\infty, 0)$. By properties of moment-generating functions, $\frac{\mathrm{d}\mathbb{E}\exp(tX_1)}{\mathrm{d}t}$ evaluated at $t = 0$ equals $\mathbb{E}X_1$, so (A.9) equals 0 at $t = 0$. It in turn suffices to show the derivative of (A.9), namely

$$\Sigma_{11} - \frac{\mathrm{d}^2 \log \mathbb{E}\exp(tX_1)}{\mathrm{d}t^2} \tag{A.10}$$

is non-negative in $t \in \mathbb{R}$.

By Tallis (1961), denoting the first column of $\boldsymbol{\Sigma}$ as $\boldsymbol{\Sigma}_1$, the moment-generating function of the marginal distribution of $X_1$ is

$$\frac{\int_{\mathbb{R}^p_+ - \boldsymbol{\mu} - t\boldsymbol{\Sigma}_1} \exp\left(-\frac{1}{2}\boldsymbol{x}^\top \boldsymbol{\Sigma}^{-1} \boldsymbol{x}\right) \mathrm{d}\boldsymbol{x}}{\int_{\mathbb{R}^p_+ - \boldsymbol{\mu}} \exp\left(-\frac{1}{2}\boldsymbol{x}^\top \boldsymbol{\Sigma}^{-1} \boldsymbol{x}\right) \mathrm{d}\boldsymbol{x}} \exp\left(t\mu_1 + \frac{1}{2}t^2 \Sigma_{11}^2\right).$$

(A.10) thus becomes

$$-\frac{\mathrm{d}^2}{\mathrm{d}t^2} \log \int_{\mathbb{R}^p_+ - \boldsymbol{\mu} - t\boldsymbol{\Sigma}_1} \exp\left(-\frac{1}{2}\boldsymbol{x}^\top \boldsymbol{\Sigma}^{-1} \boldsymbol{x}\right) \mathrm{d}\boldsymbol{x}.$$

Showing this is non-negative in $t \in \mathbb{R}$ is equivalent to showing that the integral itself is log-concave in $t$. But

$$\int_{\mathbb{R}^p_+ - \boldsymbol{\mu} - t\boldsymbol{\Sigma}_1} \exp\left(-\frac{1}{2}\boldsymbol{x}^\top \boldsymbol{\Sigma}^{-1} \boldsymbol{x}\right) \mathrm{d}\boldsymbol{x} = \int_{\mathbb{R}^p} \exp\left(-\frac{1}{2}\boldsymbol{x}^\top \boldsymbol{\Sigma}^{-1} \boldsymbol{x}\right) \mathbb{1}_{\mathbb{R}^p_+ - \boldsymbol{\mu}}(\boldsymbol{x} + t\boldsymbol{\Sigma}_1) \mathrm{d}\boldsymbol{x}$$

with $\exp\left(-\frac{1}{2}\boldsymbol{x}^\top \boldsymbol{\Sigma}^{-1} \boldsymbol{x}\right)$ log-concave in $\boldsymbol{x}$ and $\mathbb{1}_{\mathbb{R}^p_+ - \boldsymbol{\mu}}(\boldsymbol{x} + t\boldsymbol{\Sigma}_1)$ log-concave in $(\boldsymbol{x}, t)$ since $\mathbb{R}^p_+ - \boldsymbol{\mu}$ is a convex set (half-space). Here $\mathbb{1}_S(\cdot)$ is the indicator function of a set $S$. Since log-concavity is closed under multiplication and integration over $\mathbb{R}^p$, the integral is indeed log-concave, and our proof of the bound on the sub-gaussian parameter of $X_j - \mathbb{E}X_j$ is complete. The tail bound follows from 5) of Lemma A.5.

Now by 1) and 2) of Lemma A.5,

$$\|X_j\|_{\psi_2} \leq 2\sqrt{\Sigma_{jj}/e} + \mathbb{E}X_j.$$

If $h_0$ is a function bounded by $M_0$, then by definition

$$\|X_j h_0(X_k)\|_{\psi_2} \leq M_0 \left(2\sqrt{\Sigma_{jj}/e} + \mathbb{E}X_j\right).$$

By 1) and 2) of Lemma A.5 again,

$$\tau(X_j h_0(X_k) - \mathbb{E}X_j h_0(X_k)) \leq \sqrt{e}\|X_j h_0(X_k) - \mathbb{E}X_j h_0(X_k)\|_{\psi_2}$$
$$\leq 2\sqrt{e}\|X_j h_0(X_k)\|_{\psi_2}$$
$$\leq 2M_0(2\sqrt{\Sigma_{jj}} + \sqrt{e}\mathbb{E}X_j).$$

The tail bound thus follows from the first inequality using 5) of Lemma A.5. By 2),

$$\tau\left(\frac{1}{n}\sum_{i=1}^n X_j^{(i)} h_0(X_k^{(i)}) - \mathbb{E}X_j h_0(X_k)\right) \leq \frac{2M_0}{\sqrt{n}}(2\sqrt{\Sigma_{jj}} + \sqrt{e}\mathbb{E}X_j),$$

$$\left\|\frac{1}{n}\sum_{i=1}^n X_j^{(i)} h_0(X_k^{(i)}) - \mathbb{E}X_j h_0(X_k)\right\|_{\psi_2} \leq \frac{4M_0}{\sqrt{en}}(2\sqrt{\Sigma_{jj}} + \sqrt{e}\mathbb{E}X_j).$$

2. By the proof of 1) of this lemma, $\|X_j\|_{\psi_2} \leq 2\sqrt{\Sigma_{jj}/e} + \mathbb{E}X_j$, and by 3) of Lemma A.5,

$$\|X_j X_k\|_{\psi_1} \leq (2\sqrt{\Sigma_{jj}/e} + \mathbb{E}X_j)(2\sqrt{\Sigma_{kk}/e} + \mathbb{E}X_k) \leq \max_j \left(2\sqrt{\Sigma_{jj}/e} + \mathbb{E}X_j\right)^2.$$

Since $h_0$ is a function bounded by $M_0$, by definition

$$\|X_j X_k h_0(X_\ell)\|_{\psi_1} \leq M_0 \max_j \left(2\sqrt{\Sigma_{jj}/e} + \mathbb{E}X_j\right)^2.$$

Then by 1) of Lemma A.5 again,

$$\|X_j X_k h_0(X_\ell) - \mathbb{E}X_j X_k h_0(X_\ell)\|_{\psi_1} \leq 2M_0 \max_j \left(2\sqrt{\Sigma_{jj}/e} + \mathbb{E}X_j\right)^2.$$

The tail bound then follows from 6) of Lemma A.5.

□



\end{document}